\documentclass[aps,prd,twocolumn,showpacs,superscriptaddress]{revtex4}

\usepackage{amssymb}
\usepackage{amsmath}
\usepackage{amsfonts}
\usepackage{graphicx}
\usepackage{epsfig}
\usepackage{psfrag}
\usepackage{bbm}
\usepackage{color}

\begin{document}

\title{Quantum state preparation and macroscopic entanglement in gravitational-wave detectors}

\author{Helge M\"uller-Ebhardt}
\author{Henning Rehbein}
\affiliation{Max-Planck-Institut f\"ur Gravitationsphysik
(Albert-Einstein-Institut) and Universit\"at Hannover,
Callinstr.~38, 30167 Hannover, Germany}
\author{Chao Li}
\author{Yasushi Mino}
\affiliation{California Institute of Technology, M/C 130-33, Pasadena, California 91125}
\author{Kentaro Somiya}
\affiliation{Max-Planck-Institut f\"ur Gravitationsphysik
(Albert-Einstein-Institut), Am M\"uhlenberg 1, 14476 Potsdam,
Germany}
\author{Roman Schnabel}
\author{Karsten Danzmann}
\affiliation{Max-Planck-Institut f\"ur Gravitationsphysik
(Albert-Einstein-Institut) and Universit\"at Hannover,
Callinstr.~38, 30167 Hannover, Germany}
\author{Yanbei Chen}
\affiliation{California Institute of Technology, M/C 130-33,
Pasadena, California 91125} \affiliation{Max-Planck-Institut f\"ur
Gravitationsphysik (Albert-Einstein-Institut), Am M\"uhlenberg 1,
14476 Potsdam, Germany}

\date{\today}

\begin{abstract}
Long-baseline laser-interferometer gravitational-wave detectors
are operating at a factor of $\sim 10$ (in amplitude) above the
standard quantum limit (SQL) within a broad frequency band (in the
sense that $\Delta f \sim f$).  Such a low classical noise budget
has already allowed the creation of a controlled 2.7\,kg
macroscopic oscillator with an effective eigenfrequency of 150\,Hz
and an occupation number of $\sim 200$.  This result, along with
the prospect for further improvements, heralds the new possibility
of experimentally probing macroscopic quantum mechanics (MQM) ---
quantum mechanical behavior of objects in the realm of everyday
experience --- using gravitational-wave detectors.  In this paper,
we provide the mathematical foundation for the first step of a MQM
experiment: the preparation of a macroscopic test mass into a
nearly minimum-Heisenberg-limited Gaussian quantum state, which is
possible if the interferometer's classical noise beats the SQL in
a broad frequency band. Our formalism, based on Wiener filtering,
allows a straightforward conversion from the classical noise
budget of a laser interferometer, in terms of noise spectra, into
the strategy for quantum state preparation, and the quality of the
prepared state.  Using this formalism, we consider how Gaussian
entanglement can be built among two macroscopic test masses, and
the performance of the planned Advanced LIGO interferometers in
quantum-state preparation.
\end{abstract}


\maketitle

\section{Introduction}

An international array of  first-generation long-baseline laser
interferometric gravitational-wave (GW) detectors
(LIGO~\cite{SHOEMAKER2004}, VIRGO~\cite{FIORE2002},
GEO~\cite{WIL2002} and TAMA~\cite{ANDO2001}) are reaching their
design sensitivities.
\begin{figure}[t]
\includegraphics[width=\linewidth]{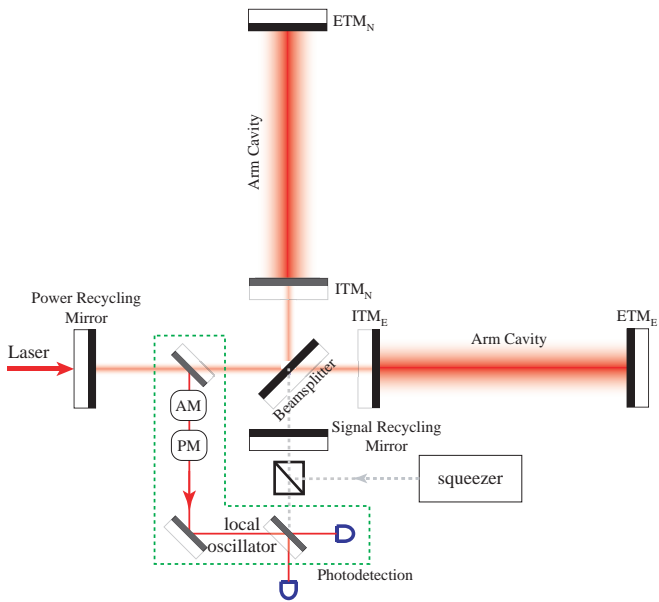}
\caption{Schematic plot of the planned Advanced
LIGO~\cite{advLIGO} interferometer: a power- and signal-recycled
Michelson interferometer with cavities in the arms and a homodyne
detection scheme at the dark port.}\label{Fig:AdvLIGO}
\end{figure}
These Michelson interferometers have been built to measure
GW-driven relative length changes (within a detection band from
10\,Hz to 10\,kHz) between the mirror-endowed test masses which
are hung as pendulums with an eigenfrequency far below the
detection band. Resonant cavities are used to enhance the
sensitivity by increasing the circulating optical power and the
signal storage time. In Michelson interferometers usually the
total differential mode of motion between the arm cavity mirrors, 
in the following always represented by the position operator
\begin{equation}
\hat x = \left(\hat x_{\rm ITM_E}-\hat x_{\rm
ETM_E}\right)-\left(\hat x_{\rm ITM_N}-\hat x_{\rm ETM_N}\right),
\end{equation}
(for the nomenclature see Fig.\,\ref{Fig:AdvLIGO}) 
is measured via a homodyne detection of the modulation fields
(also called side-band fields) leaking out at the dark port of the
interferometer. Current GW
interferometers are already quantum limited at high frequencies by
the shot noise. Next generation interferometers, such as the
planned Advanced LIGO detector~\cite{advLIGO} (cf.
Fig.\,\ref{Fig:AdvLIGO}), are expected to be quantum noise limited
at nearly all frequencies in the detection band. Advanced LIGO
will therefore operate at its free mass {\it standard quantum
limit} (SQL)~\cite{CTDSZ1980,BrKh1999a,Bra1968} at which the
back-action noise is comparable to the shot noise level. The
position-referred spectral density of the SQL at the sideband frequency 
$\Omega$ is given by
\begin{equation}\label{Eq:SQL} 
S_{\rm SQL} (\Omega) \equiv \frac{2\,\hbar}{m\,\Omega^2}\,.
\end{equation}
The SQL is the minimum noise spectrum achievable by a linear
quantum measurement of the position with uncorrelated sensing and
back-action noise. Here $m$ is the reduced mass of all arm cavity
mirrors, or $1/4$ of the individual mirror mass.

Improvement of sensitivities to GWs beyond Advanced LIGO will
require to surpass the SQL significantly in a broad frequency
band.  Various  conceptual strategies exist for building
interferometers with broadband sub-SQL quantum
noise~\cite{KLMTV2001,Cou2002,Kha2002,Pu2002,PuCh2002,Che2003,HCCFVDS2003,Kha2003,Dan2004,BuCh2004,Cor2004,Dan2005,Kha2006,Kha2007,RMSLSDC2007,RMSDSDC2008},
while there is also much effort devoted to lowering classical
noise below the SQL. For example, (i) the CLIO interferometer is
currently being cooled down to a few tens of Kelvin, and has a
theoretical thermal noise budget below the
SQL~\cite{MUYHKIOKT2004}, (ii) non-spherical mirrors are being 
developed that support non-Gaussian modes
which average better over mirror surfaces and are thus less
susceptible to thermal noise~\cite{BBFDS2006,Shaugh2006,BoTh2006},
(iii) coating structures are also being optimized for coating
thermal noise~\cite{Evans2008,Kimble2008}.

This paper, however, is devoted not to the improvement of the
detector's sensitivity to GWs, but to the study of quantum
mechanical behavior of its test masses. Recently, it has been
reported by the LIGO Scientific Collaboration in
Ref.~\cite{LSC2008} that a certain (controlled) mode of the
mirror's differential motion in the LIGO detector located in
Hanford was experimentally cooled down to $1.4\,\mu$K, with an
effective occupation number of around 200. Furthermore, there is a
number of other experiments considering different smaller-scale
mechanical
structures~\cite{CHP1999,HJHS2003,KlBo2006,CWBOSSWM2007,Gossi08,Vinante2008}.
The goal of such {\it cold damping}~\cite{MVT1998} experiments is
to reach the oscillator's ground state. As a real pure quantum
state is approached, the semiclassical model as used in the above
references will certainly break down and the quantum noise effects
in the measurement process have to be included.

One aim of this paper is at providing a mathematical foundation
guiding future experimental efforts of quantum-state preparation
--- one that is straightforward to apply to experimental
situations. Note that our general formulation applies not only to
gravitational-wave detectors but to the whole bunch of experiments
related to quantum state preparation. Quantum mechanically, we
consider a problem in which an object is being continuously
measured by the optical field, while it is simultaneously subject
to noisy forces. For such problems, the stochastic master equation
(SME)~\cite{Gar2004} is a readily available tool to simulate
quantum-state preparation in quantum mechanical systems; a Riccati
equation associated with the SME provides the remaining
uncertainty of the object, when all measurement data are taken
into account. Formally, this approach only treats Markovian
systems (since only Markovian systems allow us to project the
optical field being measured and trace off the noise fields at
every infinitesimal time step, resulting in a closed-form
evolution of the test-mass density operator), while
non-Markovianity is prevalent in experiments such as in GW detectors: 
virtually none of the noise sources are white, and we have the added complexity that
the cavity mode often couples strongly with the test masses, i.e.
it cannot be {\it adiabatically eliminated}~\cite{Gar2004}, and
must be evolved together with the test masses.  Since we only
consider linear systems with Gaussian noise, and we only care
about the test masses' state after the initial transient has died
down, the SME and the Riccati equation, which also characterizes
the exact way of the transient decays, are not entirely necessary.
Instead, we have found that a Wiener filtering approach, in which
the measurement data is filtered with the optimal causal Wiener
filter to obtain instantaneous optimal estimates for position and
momentum of the test masses~\cite{MRSDC2007}, suffices, and is
most straightforwardly connected with experimental calibration of
the system.  An example of the power of the Wiener filtering
approach has already been demonstrated in
Ref.~\cite{CDMMRSS2008b}, where an optimal controller that yields
a steady quantum state with minimum uncertainty (or von Neumann
entropy) has been derived for a general linear (Markovian as well
as non-Markovian) quantum measurement process.

A direct application of the Wiener filtering approach is to
explore how quantum a macroscopic test mass can be prepared in a certain
environment, which is the second aim of this paper. We measure the
{\it purity} of the quantum state of an individual test mass (or a
single mechanical degree of freedom) through the Heisenberg
uncertainty product or the following quantity
\begin{equation} \label{Eq:Uquant}
U\equiv \frac{2}{\hbar}\sqrt{V_{xx}\,V_{pp}-V_{xp}^2}\,,
\end{equation}
which is unity for a pure state. This quantity can also be
converted into an effective occupation number. Here $V_{xx}$,
$V_{pp}$ and $V_{xp}$ are the second-order moments of position and
momentum of the Gaussian state. We will show that a low classical
noise budget which is completely below the SQL for a broad
frequency band allows the quantity $U$ to become close to unity.
In particular for simple systems with a total classical noise
spectrum $S_{\rm cl}(\Omega)$ which is dominated by a white
sensing noise and a white force, we have obtained the simple
relation
\begin{equation} \label{Eq:Uapprox}
U \approx 1 + \min_{\Omega} \left\{\frac{S_{\rm
cl}(\Omega)}{S_{\rm SQL}(\Omega)}\right\}\,.
\end{equation}
But we will also explore how a realistic noise budget for the
planned Advanced LIGO detector
--- as well as an extension of Advanced LIGO with plausible
improvement --- can best be taken advantage of through an
optimized optical configuration that minimizes $U$. When two
independent mechanical degrees of freedom are considered, the
formalism, which we present in this paper, has already been
applied to show that the production of quantum entanglement
between the macroscopic end mirrors is possible for sub-SQL laser
interferometers~\cite{MRSDC2007}.

An experiment testing MQM should be divided into different stages 
which are separated in time: a preparation stage, where the test mass 
will be continuously observed; an optional free-evolution stage; and a 
verification stage. One will need to collect statistics from a huge number 
of identical trials as it is required from quantum mechanics.
This present paper is the first one of a paper-series: this one
deals with the preparation of macroscopic conditional quantum
states; while a second paper~\cite{CDMMRSS2008} will study the
verification of such macroscopic quantum states. 

This paper is organized as follows: in Sec.\,\ref{Sec:Wiener} we will briefly
review the theoretical basics of Wiener filtering. In
Sec.\,\ref{Sec:Marko} we will study analytically the conditional
variances of the simplified model using only Markovian dynamics.
We will introduce a flexible homodyne detection angle and input
squeezing. In Sec.\,\ref{Sec:MacroEnt} we will extend the analysis
done in Ref.~\cite{MRSDC2007} about macroscopic entanglement. In
Sec.\,\ref{Sec:NonMarko} we will study test-masses in a cavity
with finite bandwidth and we will treat more realistic,
non-Markovian noise sources. These preliminary studies result in
an investigation of quantum-state preparation in Advanced LIGO in
Sec.\,\ref{Sec:AdvLIGO}. Finally, in Sec.\,\ref{Sec:Con} we will
summarize our main conclusions.

\section{Wiener filtering} \label{Sec:Wiener}

For systems under continuous measurement, the conventional
approach is to describe the joint system-measurement-data
evolution using a {\it stochastic master equation}
(SME)~\cite{Mil1996,Gar2004,DTPW1999,HJHS2003}, which is a set of
stochastic differential equations that simulates the joint
evolution of the system's {\it conditional density matrix}
$\hat\rho^{\rm cond}$ and measurement data $y(t)$.  As the
simplest example, for a harmonic oscillator with position $\hat x$
being measured continuously by a Markovian measurement device (one
that has uncorrelated measurement noise at different times and
constant measurement strength), the SME reads,
\begin{align}
{\rm d}\hat\rho^{\rm cond} &= -\frac{\rm i}{\hbar}\, [H,\,\hat
\rho^{\rm cond}]\, {\rm d}t - \frac{\alpha^2}{4 \hbar^2}\, \left[
\hat x,\, \left[\hat x,\, \hat \rho^{\rm cond}\right]\right]\,
\mathrm{d} t \nonumber
\\ &+ \frac{\alpha}{\sqrt{2} \hbar}\, \left(\left\{\hat x,\, \hat \rho^{\rm cond}\right\}
- 2\, \langle \hat x \rangle\, \hat \rho^{\rm cond}\right)\,
\mathrm{d} W\,, \\
{\rm d} \hat y &= \frac{\alpha}{\hbar}\, \langle \hat x \rangle \
\mathrm{d} t + \mathrm{d} W/\sqrt{2}\,,
\end{align}
with a coupling constant $\alpha$. Here the {\it conditional
quantum state}~\cite{Barchielli1992} is defined as the projection
of the joint system-device quantum state to the sub-space in which
the readout observable $\hat y$ has definite values of $\hat y(t')
= y(t')\ \forall\, 0<t'<t$. The Wiener increment $dW$ describes a
stochastic process that simultaneously drives the conditional
quantum state and the measurement data, where both are stochastic
processes. Different realizations of $dW$ correspond to different
possible scenarios that could take place in reality.

In practice, it is not enough to only describe the stochastic
process, we need to be able to obtain the conditional quantum
state at any given time $t$, based on the system's initial quantum
state $\hat{\rho}^{\rm cond}(0)$ and measurement results
$\{y(t'),\,0<t'<t\}$.  This corresponds to a {\it filtering
problem} in classical stochastic calculus.  The probability
distribution of any state variable $\hat x$ is simply the
conditional probability
\begin{equation}
P[\hat x(t)\,|\,\{y(t'),\, 0<t'<t\}]\,,
\end{equation}
while the conditional expectation of $\hat x$ can be written as a
functional over $\{y(t'),\, 0<t'<t\}$,
\begin{equation}
x^{\rm cond} (t) = E[\hat x(t)\,|\,\{y(t'),\, 0<t'<t\}]\,.
\end{equation}
For linear systems with Gaussian states only expectations of
quantities linear and quadratic in the variables are needed. The
former can be obtained through a linear filter over $\hat y$,
while the latter can be obtained by solving a time-domain Riccati
equation. If the measurement process has started for sufficiently
long time --- much longer than the time constant of transients ---
then the filters over $\hat y$ as well as the second-order moments
are stationary. They can be obtained through the theory of Wiener
filtering~\cite{Wiener1949}. This will be the situation that we
will consider in this paper.

Although the most general theory of quantum filtering is quite
distinct from classical filtering~\cite{BvHJ07},  we will simplify
the situation through two steps, and return to an essentially
classical filtering problem. First, let us construct a model of
the quantum measurement process. Let $\hat{y}(t)$ be the {\it
Heisenberg operator} of the measurement output (e.g., a particular
quadrature of the out-going optical field) and $\hat{s}(t)$ any
system observable. Then the principle of simultaneous
measurability, i.e. the observable $\hat y(t')$ at different times
can be measured individually to arbitrary accuracy without
imposing any fundamental limits, and the principle of causality,
i.e. measurement observable at present does not respond to future
forces to be exerted onto the system, will dictate
that~\cite{BuCh2002}
\begin{align}
0 &= [\hat y(t),\, \hat y(t')] \quad \forall t,t'\,,  \label{Eq:comyy} \\
0 &= [\hat o(t),\, \hat y(t')] \quad \forall t>t'\,.
\label{Eq:comxy}
\end{align}
Note here that $\hat o$ could be any any state variable such as
momentum and position, or even the density operator of the system
being measured.  This means that any filtering of the operator
$\hat y(t')$ for $0<t'<t$ can be considered as a classical
process, i.e. treated with classical linear control theory, as
long as the state of the system at $t$, or $\hat\rho(t)$ is
considered. Here we write formally that
\begin{eqnarray}
&&\hat \rho^{\rm cond}_{\{y(t'),\, 0<t'<t\}} \nonumber\\
&& \quad = \frac{\mathcal{P}_{[\hat y (t') = y (t'),\, t' < t]} \
\hat \rho \ \mathcal{P}_{[\hat y (t') = y (t'),\,t' <
t]}}{\mathrm{tr}\left[\mathcal{P}_{[\hat y (t') = y (t'),\, t' <
t]} \ \hat \rho \ \mathcal{P}_{[\hat y (t') = y (t'),\,t' <
t]}\right]}\,,
\end{eqnarray}
where $\mathcal{P}$ projects onto the subspace on which the
measurement operator takes the measured value. Henceforth the
dependence of $\hat \rho^{\rm cond}$ on $\{y(t'), t'<t\}$ will not
be written explicitly, as it is done in most of the literature.

Second, we restrict ourselves to {\it stable} linear systems with
Gaussian states and to later times when the initial states are no
longer important. In this case only first- and second-order
moments in linear observables need to be considered. We also
assume all linear observables to have zero unconditional
expectation values. Suppose we have observables $\hat x_l$ with
$l=1,2,\ldots n$, and proceed as in classical Wiener filtering.
The key is to decompose the operator $\hat x_l$ into
\begin{equation}
\label{xldecompose} \hat{x}_l(t) =\int^{t}_{-\infty} {\rm d}
t'\,K_l(t-t') \hat y(t') + \hat{R}_l(t)
\end{equation}
with
\begin{equation}
\label{Ryindependence} \langle \hat R_l(t)\, \hat y(t') \rangle =
\mathrm{tr} \left[\hat\rho\, \hat R_l(t)\, \hat y(t')\right]=0\,.
\end{equation}
If this can indeed be done, it can then be shown that
\begin{equation} \label{xlc}
x_{l}^{\rm cond} (t)  = \mathrm{tr} \left[ \hat\rho^{\rm cond}
\hat x_l(t)\right] = \int^{t}_{-\infty} {\rm d}t'\, K_l(t-t')\,
y(t')
\end{equation}
and
\begin{eqnarray}
\label{Vlmc} V_{lm}^{\rm cond} &\equiv& \langle \hat x_l (t) \hat
x_m (t ) \rangle^{\rm cond}_{\rm sym}
-  \langle x_l (t)  \rangle^{\rm cond} \langle x_m (t) \rangle^{\rm cond} \nonumber \\
&= & {\langle \hat R_l(t) \hat  R_m(t)\rangle}_{\rm sym}\,.
\end{eqnarray}
In this way, the conditional expectations of the linear variables
are given as linear functionals of past measurement data
$\{y(t'),\, t'<t\}$, and the conditional variances as steady state
constants. Please see Appendix\,\ref{Sec:QJ} for a more rigorous
justification.

We will now try to obtain $K_l$ and $V_{lm}^{\rm cond}$ in terms
of unconditional correlation functions, or cross spectra among
system observables $\hat x_l$ and the output $\hat y$.
Eq.~(\ref{Ryindependence}) leads to the Wiener-Hopf equation
\begin{align} \label{Eq:WienerHopf}
C_{x_l y} (t-t'')-\int_{-\infty}^t {\rm d} t'\,K_{l} (t-t') \,
C_{yy} (t'-t'') &= 0 \nonumber \\ & \forall t'' \le t\,.
\end{align}
Here
\begin{eqnarray}
C_{ab} (t-t') &\equiv& \langle \hat a (t)\,\hat b (t')
\rangle_{\rm
sym}  \nonumber \\
&= &\mathrm{tr}\left[ \hat\rho \ \frac{\hat a(t) \hat b(t') + \hat
b(t') \hat a(t) }{2}\right]
\end{eqnarray}
stands for the symmetrized time-domain two-point correlation
function between two arbitrary Heisenberg operators $\hat a (t)$
and $\hat b (t)$.

If we suppose that $K_l (t) = 0$ for $t<0$, i.e. we make sure that
the filter is a causal function, we can rewrite
Eq.\,(\ref{Eq:WienerHopf}) as
\begin{equation} \label{Eq:WienerHopf2}
C_{x_l y} (t) - \int_{-\infty}^\infty {\rm d} t'\,K_{l} (t')\,
C_{yy} (t-t') = 0 \quad \forall t \ge 0\,.
\end{equation}
In Fourier domain the condition in Eq.\,(\ref{Eq:WienerHopf2}) is
satisfied if the function
\begin{equation}
L(\Omega) = S_{x_l y} (\Omega) - K_{l} (\Omega)\, S_{yy} (\Omega)
\end{equation}
is analytic in the lower-half complex plane while the Fourier
transform of the filter function $K_l (\Omega)$ is analytic in the
upper-half complex plane. Furthermore, $L(\Omega)$ has to vanish
at infinity, because Eq.\,(\ref{Eq:WienerHopf2}) must also be
valid at $t=0$. These conditions uniquely define $K_l (\Omega)$.
We have denoted with $S_{ab} (\Omega)$ the single-sided (cross-)
spectral density among the operators $\hat a$ and $\hat b$,
related to the two-point correlation as
\begin{equation}
C_{ab} (t) = \frac{1}{2}\int_{-\infty}^{+\infty}
\frac{d\Omega}{2\pi }\, S_{ab}(\Omega)\, e^{-i\Omega t}\,.
\end{equation}
The  Wiener-Hopf method provides the solution for $K_l (\Omega)$
as
\begin{equation}
\label{Kl} K_l(\Omega) = \frac{1}{s_{y}^+ (\Omega)}
\left[\frac{S_{x_ly} (\Omega)}{s_{y}^- (\Omega)}\right]_+\,,
\end{equation}
where we have split $S_{yy} (\Omega) = s_{y}^+ (\Omega)\, s_{y}^-
(\Omega)$ in such a way that $s_{y}^+$ ($s_{y}^-$) and its inverse
are analytic functions in the upper-half (lower-half) complex
plane. Because $S_{yy}(\Omega) $ is in general a rational function
of $\Omega^2$ with real coefficients, we expect that
$s_y^+(\Omega) =[ s_y^-(\Omega^*)]^*$. In addition,
$[F(\Omega)\ldots]_+$ stands for taking the component of a
function whose inverse Fourier transform has support only in
positive times. Operationally, this could be obtained either by
decomposing $F(\Omega)$ into
\begin{equation}
F(\Omega) = \sum_k \frac{\alpha_k}{\Omega-\Omega_k}
\end{equation}
and only keeping terms whose $\Omega_k$ has negative imaginary
parts, or by inverse Fourier transforming $F(\Omega)$ into the
time domain, eliminate the positive-time component, and then
Fourier transform back. [Note that both approaches will become
ambiguous when $F(\Omega)$ does not approach zero when
$\Omega\rightarrow +\infty$.] Inserting Eq.~(\ref{Kl}) into
Eqs.~(\ref{xldecompose}) and (\ref{Vlmc}), we obtain
\begin{eqnarray} \label{Eq:Wcondcov}
V_{lm}^{\rm cond} =  \int_0^\infty \frac{{\rm
d}\Omega}{2\pi}\left(S_{x_l x_m}- \left[\frac{S_{x_l
y}}{s_{y}^-}\right]_+\left[\frac{S_{x_my}}{s_{y}^-}\right]_+^*\right)\,.
\label{Eq:Wcondxx} \end{eqnarray}

A more transparent understanding of the filtering problem can be
obtained when we formally define the {\it causally whitened
output} as
\begin{equation}
\hat z (\Omega) \equiv \frac{1}{s_y^+(\Omega)}\ \hat y(\Omega)\,.
\end{equation}
Because  $1/s_y^+$ is analytic in the upper-half complex plane,
$\hat z$ can be written as an integral over the history of $\hat
y$. The random process $\hat z$ has a white spectrum. Moreover, we
can write
\begin{equation}
K_l (\Omega)\, \hat y(\Omega) =  [S_{x_l z}(\Omega) ]_+\, \hat
z(\Omega)\,.
\end{equation}
In other words,
\begin{equation}
\label{Klsimp} \langle x_l(t)\rangle^{\rm cond} =
\int_{-\infty}^{t} {\rm d} t'\, C_{x_l z}(t-t') z(t')
\end{equation} and
\begin{equation}
\label{Vlmsimp} V_{lm}^{\rm cond} = V_{lm} - \int_{-\infty}^0 {\rm
d} t'\, C_{x_l z}(-t') C_{x_m z}(-t')\,.
\end{equation}
Clearly, Eqs.~(\ref{Klsimp}) and (\ref{Vlmsimp}) are simply
continuous versions of linear regression over a set of independent
random variables.

In this paper with tools of Wiener filtering, we no longer need to
write down SMEs. In fact, for realistic systems with multiple
colored noise sources and non-Markovian dynamics, SMEs can only be
obtained by increasing the dimension of the problem, which will
definitely become cumbersome.

\section{Measurement configurations with Markovian noise}
\label{Sec:Marko}

\subsection{General discussion}

Let us start our discussion generally with an abstract continuous
linear Markovian measurement process, which monitors the center of mass 
position of a simple harmonic oscillator. The Heisenberg equations of
motion in the frequency domain can be written as
\begin{eqnarray}
\hat y &=& \hat Z + \hat x\,, \label{Eq:simpeomy} \\
\hat x &=& R_{xx} (\Omega) \ \hat F\,, \label{Eq:simpeomx}
\end{eqnarray}
where the two noise operators $\hat Z$ and $\hat F$ both have a
white spectrum. In the time domain, $\hat Z$ and $\hat F$ are
white noise, with two-time correlation functions proportional to
delta function.  Such statistical characteristics make the
measurement process a Markovian one. Thus, we assume that $\hat Z$
and $\hat F$ have the single-sided (cross-) spectral densities
$S_{ZZ}\geq 0$, $S_{FF}\geq 0$ and $S_{ZF} \in \mathbb{R}$, which
satisfy the Heisenberg relation of the measurement
process~\cite{BrKh1999a}
\begin{equation} \label{Eq:measurepurity}
S_{ZZ}\, S_{FF} - S_{ZF}^2 = \mu \hbar^2\,,\qquad \mu \ge 1
\end{equation}
which arises from the requirement that the Heisenberg operators of
the out-going field  at different times must commute, and
guarantees that the level of back action is just enough to enforce
the Heisenberg uncertainty relation of the test
mass~\cite{BrKh1999a}. In the case of Gaussian noise only, we have
$\mu=1$ if and only if the measurement process is purely quantum.

The linear response function of a damped harmonic oscillator is
given by
\begin{equation}
R_{xx} (\Omega) = -\frac{1}{m\,\left(\Omega^2 + {\rm
i}\,\gamma_m\,\Omega-\omega_m^2\right)}\,,
\end{equation}
with the eigenfrequency $\omega_m$, the damping rate $\gamma_m \ll
\omega_m$ and the mass $m$. Then we can easily assemble the
single-sided spectral densities of the measurement process as
\begin{align}
S_{yy} (\Omega) &= m^2|R_{xx}(\Omega)|^2 \left(\Omega^4 -
\Omega^2(2q_1 - \gamma_m^2) + q_2^2\right) S_{ZZ}\,, \label{Eq:SYY} \\
S_{xy} (\Omega) &= - m|R_{xx}(\Omega)|^2 \left(\!\Omega ^2-{\rm
i}\gamma _m \Omega -\omega _m^2-\frac{S_{FF}}{S_{ZF}}\right)
S_{ZF}\,, \label{Eq:SXY} \\
S_{xx} (\Omega) &= |R_{xx}(\Omega)|^2 S_{FF}\,, \label{Eq:SXX}
\end{align}
where we have defined the coefficients
\begin{align}
q_1 &\equiv \omega_m^2+\frac{S_{ZF}}{m\, S_{ZZ}}\,, \label{Eq:q1def} \\
q_2 &\equiv \sqrt{\omega_m^4 + 2\,\omega_m^2\
\frac{S_{ZF}}{m\,S_{ZZ}} + \frac{S_{FF}}{m^2 S_{ZZ}}}\,.
\label{Eq:q2def}
\end{align}
Both, $q_{1,2}$, have the dimension of frequency squared, while
$q_2$ is always positive. We also note that $|q_1| \le q_2$.

From Eq.\,\eqref{Eq:SYY} we can recover a quantum limit of the
measurement process: if we have $S_{ZF} = 0$, the spectral density
of the measurement noise satisfies
\begin{eqnarray} S_{yy}
(\Omega) \geq 2\,\sqrt{S_{ZZ}\,S_{FF}}\,|R_{xx}(\Omega)|
&\geq&  2\, \sqrt{\mu}\,\hbar\,|R_{xx}(\Omega)| \nonumber \\
&=& \sqrt{\mu}\, S_{\rm SQL} (\Omega)\,. \label{Eq:SQL2}
\end{eqnarray}
For a free mass we can then recover Eq.\,\eqref{Eq:SQL} from
Eq.\,\eqref{Eq:SQL2}. Note that for the free mass the first
inequality sign in Eq.\,\eqref{Eq:SQL2} becomes an equality sign
at $\Omega = \Omega_q$, where the measurement frequency is defined
by
\begin{equation} \label{Eq:touchingfreq}
\Omega_q^2 \equiv \sqrt{\frac{S_{FF}}{m^2\,S_{ZZ}}}\,.
\end{equation}
Therefore, the measurement frequency $\Omega_q$ is the frequency
at which the noise spectral density of a Markovian quantum
measurement process (with $S_{FZ}=0$) approaches most its free
mass SQL.

Now it is straightforward to derive the conditional variances
assuming $\hat p = - {\rm i}\,m\, \Omega\, \hat x$ and using the
Wiener filtering method, i.e. Eq.\,\eqref{Eq:Wcondcov}. For this
it has actually been crucial to first spectral factorize $S_{yy}
(\Omega)$. In the perfect oscillator limit with $\gamma_m
\rightarrow 0$ the conditional {\it covariance matrix} can be put
into the following form
\renewcommand{\arraystretch}{2.5}
\begin{equation} \label{Eq:condcov}
\mathbf{V} = {\sqrt{\mu}} \
\mathbf{D} \ \left(\begin{array}{cc}
\displaystyle \sqrt{\frac{2q_2}{q_1+q_2}} &  \displaystyle \sqrt{\frac{q_2-q_1}{q_2+q_1}} \\
\displaystyle \sqrt{\frac{q_2-q_1}{q_2+q_1}} &   \displaystyle \sqrt{\frac{2q_2}{q_1+q_2}}
\end{array}\right) \ \mathbf{D}^T\,,
\end{equation}
where we have defined
\begin{equation}
\mathbf{D} = \left(\begin{array}{cc}
\displaystyle \sqrt{\frac{\hbar}{{2 m\sqrt{q_2}}}} & 0\\
0 & \displaystyle {\sqrt{\frac{\hbar m\sqrt{q_2}}{2}}}
\end{array}\right)\,.\end{equation}
Eq.\,\eqref{Eq:condcov} gives us the most general covariance
matrix of the conditional Gaussian state of a lossless harmonic
oscillator under any linear Markovian position measurement. In the
covariance matrix $\mathbf{V}$, the matrix $\mathbf{D}$ sets the
scale of the quantum fluctuations to be comparable to those of the
vacuum state of a harmonic oscillator with eigenfrequency
$\sqrt{q_2}$. Depending on the ratio $q_1/q_2$, which always lies
between $-1$ and $+1$, the noise ellipse of the vacuum state is
deformed into one where position and momentum are correlated
(unless if $q_1/q_2 \approx 1$) while the area is conserved. This
corresponds to a unitary transformation among Gaussian states.
Then $\mu$ finally enlarges the noise ellipse with a uniform
factor, converting the pure state into a mixed state (unless if
$\mu=1$). We have found that the conditional variances completely
coincide with those obtained from SMEs. With
Eq.\,\eqref{Eq:condcov} we can for instance easily reproduce
Eqs.\,(2.8a)--(2.8c) from Ref.~\cite{DTPW1999}.

A Gaussian state is pure if and only if its uncertainty product is
Heisenberg-limited. Therefore, it makes sense to quantify the
purity of the conditional state by its uncertainty product, here
given by
\begin{equation} \label{Eq:conddet} \det\mathbf{V} =
V_{xx} V_{pp} -V_{xp}^2 = \mu\, \frac{\hbar^2}{4}\,,
\end{equation}
which is identical to the purity of the measurement process. This
simply shows that in the Markovian case, any measurement will
produce a pure conditional state of a lossless harmonic oscillator
if and only if it is a quantum measurement. Moreover, the
uncertainty product is even independent of the system's mechanical
properties such as the oscillator's mass and eigenfrequency. Note
that in the Appendix\,\ref{Sec:Neff} we have introduced how the
uncertainty product is related to an effective occupation number
$\mathcal{N}_{\rm eff}$. It turns out, when using
Eq.\,\eqref{Eq:eigeneff}, that here the effective eigenfrequency
for the effective occupation number is given by $\omega_{\rm eff}
= \sqrt{q_2}$.

The covariance matrix in Eq.\,\eqref{Eq:condcov} becomes obviously
diagonal and the correlation between $\hat x$ and $\hat p$ in the
conditional state vanishes if and only if $q_1 = q_2$. But this is
strictly forbidden due to the Heisenberg uncertainty principle.
However, in a certain limit they can become very
close~\cite{CDMMRSS2008b}. With a higher difference in $q_1$ and
$q_2$ the correlation in $\hat x$ and $\hat p$ increases --- and
in turn also the uncertainty product.

In order to obtain the conditional state, as given in
Eq.\,\eqref{Eq:condcov}, the measurement data has to be filtered
in real time using the Wiener filter functions for position and
momentum which are given in the frequency domain by
\begin{align}
K_x (\Omega) &= \sqrt{2}\,{\rm i}\, \sqrt{q_2-q_1}\, \frac{\Omega
-\Omega _3}{\left(\Omega -\Omega _1\right)
\left(\Omega -\Omega _2\right)}\,, \label{Eq:explwienerpx} \\
K_p (\Omega) &= {\rm i}\, m\, \left(q_2-\omega _m^2\right)\,
\frac{\Omega +\omega_m^2/\Omega _3}{\left(\Omega -\Omega _1\right)
\left(\Omega -\Omega _2\right)}\,, \label{Eq:explwienerp}
\end{align}
where $\Omega_{1,2} = (\pm \sqrt{q_2 + q_1} - {\rm i}\, \sqrt{q_2
- q_1})/\sqrt{2}$ and $\Omega_3 = {\rm i}/\sqrt{2}\, (\omega
_m^2-q_2)/\sqrt{q_2-q_1}$. Note that the poles of the Wiener
filter $\Omega_{1,2}$ are actually equal to the zeros of the
measurement's output spectrum $S_{yy} (\Omega)$, which in turn
correspond to the frequencies of maximal sensitivity and are
therefore easy to find.

Let us now have a closer look on the noise model we will use
throughout this section.
\begin{figure}[t]
\centerline{\includegraphics[height=\linewidth,angle=-90]{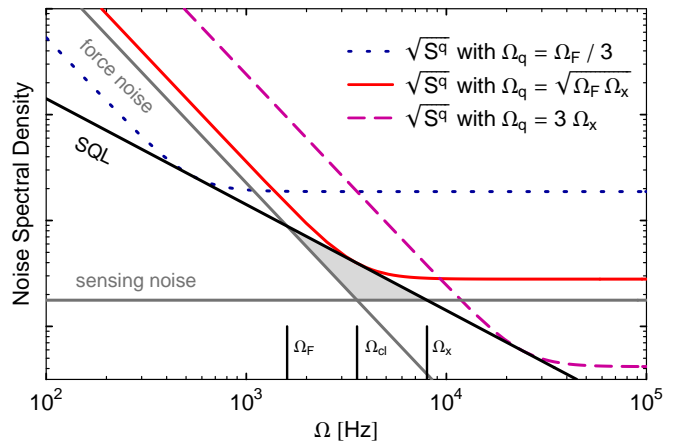}}
\caption{Example noise spectral densities (in arbitrary units) of
a Markovian measurement process observing a free mass: the quantum
noise spectral density at different values of the measurement
frequency $\Omega_q$ as well as the spectral densities
corresponding to a classical force noise (gray line) and a sensing
noise (gray line) are marked in the plot. The gray shadowed region
marks the classical noise SQL beating which we have chosen to have
$\Omega_x/\Omega_F = 5$.} \label{Fig:SNDMARKOV}
\end{figure}
We will only consider quantum measurement processes and categorize
the noise into two groups: {\it (i)} the one which is a result of
the measurement process itself will be denoted by {\it quantum
noise}; and {\it (ii)} the additional noise will be called {\it
classical noise} which does not directly arise from the
measurement process, usually has no correlation in $\hat Z$ and
$\hat F$ does not have to satisfy Eq.\,\eqref{Eq:measurepurity}.
Then the two noise sources combine as $S_{FF} = S_{FF}^{q} +
S_{FF}^{\rm cl}$ and $S_{ZZ} = S_{ZZ}^{q} + S_{ZZ}^{\rm cl}$.

The quantum noise is dominated at high frequencies by shot noise
which is covered by $S_{ZZ}^{q}$ and at low frequencies by
back-action noise which is covered by $S_{FF}^{q}$. The latter one
is represented by the radiation-pressure noise in the case of a
measurement with light. If both are uncorrelated, i.e. $S_{ZF}^{q}
= 0$, they result in the SQL. Then the quantum noise spectral
density is limited from below by the free-mass SQL as shown in
Eq.\,\eqref{Eq:SQL2} and also in Fig.\,\ref{Fig:SNDMARKOV}. The
quantum noise touches the free-mass SQL at the frequency $\Omega =
\Omega_q$.

The other noise source can also be divided into two parts: a {\it
classical force noise} $S_{FF}^{\rm cl}$ is added to $S_{FF}$
which acts directly on the center of mass of the measured object,
and is in real interferometric experiments due to for instance
seismic noise or thermal noise in the suspension of the mirrors.
The {\it classical sensing noise} $S_{ZZ}^{\rm cl}$ is only a
pseudo motion of the measured object and may be due to the
following reasons: {\it (i)} on the one hand due to thermal
fluctuations of the mirror's shape as for example mirror internal
thermal noise which makes only the mirror surface move with
respect to its center of mass; or {\it (ii)} on the other hand be
due to optical losses; or {\it (iii)} due to photo-detection
inefficiency. Therefore, our sensing noise is somehow generalized
from what is conventionally understood when using the term
``sensing noise''. Note that for detecting gravitational waves,
only the {\it total noise} matters, yet for studying quantum-state
preparation, it is important to make distinctions between sensing
and force noise, and between quantum and classical noise.
Throughout this section we will assume both classical noise
sources to have a white spectrum. Then they can be characterized
by the frequencies, $\Omega_F$ for the force noise and $\Omega_x$
for the sensing noise, at which their noise spectral density
intersects with the free-mass SQL (cf. Fig.\,\ref{Fig:SNDMARKOV}).
These classical noise frequencies are defined by the following
relations
\begin{align}
S_{FF}^{\rm cl} &= 2\,\hbar\,m\,\Omega_F^2\,, \\
S_{ZZ}^{\rm cl} &= \frac{2\,\hbar}{m\,\Omega_x^2}\,.
\end{align}
Note that the classical force noise together with the classical
sensing noise can open a window in which both are below the
free-mass SQL as indicated by the gray-shadowed region in
Fig.\,\ref{Fig:SNDMARKOV}. The total classical noise is then even
completely below the SQL in a certain frequency band, if and only
if the classical noise sources satisfy
\begin{equation} \label{Eq:clasconst}
S_{FF}^{\rm cl}\,S_{ZZ}^{\rm cl}<\hbar^2 \quad \Leftrightarrow
\quad \Omega_x/\Omega_F > 2\,,
\end{equation}
which turns into a constraint for the classical noise frequency
ratio. If Eq.\,\eqref{Eq:clasconst} holds, the classical noise is
equal to a minimum of $(2\,\Omega_F/\Omega_x)$ times the free-mass
SQL at the frequency $\Omega = \Omega_{\rm cl} \equiv
\sqrt{\Omega_F\,\Omega_x}$ which reads
\begin{equation} \label{Eq:SQLbeating}
\min_{\Omega} \left\{\frac{S_{\rm cl}(\Omega)}{S_{\rm
SQL}(\Omega)}\right\} = \frac{S_{\rm cl}(\Omega_{\rm cl})}{S_{\rm
SQL}(\Omega_{\rm cl})} = \frac{2\,\Omega_F}{\Omega_x}\,.
\end{equation}
Since here the classical noise has the largest separation to the
SQL, we can understand the factor $(2\,\Omega_F/\Omega_x)$ as the
{\it classical-noise-SQL-beating factor}.

\subsection{Very low finesse cavity and free mass scenario
with vacuum input}

Let us consider the simple situation of a laser beam incident on a
suspended mirror, where the output field is monitored by a perfect
balanced homodyne detection at a frequency-independent angle
$\zeta$. This corresponds to a mirror in a cavity with infinitely
large bandwidth (or at least much larger than $\Omega_q$, as we
shall quantify in Sec.\,\ref{SubSec:FinCav}), in which case the
dynamics of the cavity mode can be ignored or {\it adiabatically
eliminated}. Note that the following analysis is also valid for
the dark port fields entering and leaving an equal-arm Michelson
interferometer with movable end mirrors and the differential
motion between these mirrors --- but then the mirror mass $m$ in
the following discussion has to be substituted by the effective
mass $m/2$. Then the quantum measurement process can be described
by the following spectral densities~\cite{MRSDC2007}
\begin{align}
S_{ZZ} &= \frac{\hbar ^2}{\alpha ^2}\,\tan^2 \zeta+\frac{\hbar
^2}{\alpha ^2}+\frac{2\, \hbar }{m\, \Omega _x^2}\,, \\
S_{FF} &= \alpha ^2+2\, m\, \hbar\,  \Omega _F^2\,, \\
S_{ZF} &= \hbar\,\tan \zeta\,,
\end{align}
where the coupling constant is defined by
\begin{equation} \label{Eq:alpha}
\alpha = \sqrt{\frac{8\, P\, \omega_0\, \hbar}{c^2}}\,.
\end{equation}
Here $P$ is the circulating laser power and $\omega_0$ the laser
angular frequency.

In the free-mass limit, i.e. with $\omega_m \rightarrow 0$, the
conditional variances simplify to
\begin{align}
V_{xx} &= \frac{\hbar}{\sqrt{2} m \Omega _q}\sqrt{1+\tan^2 \zeta +
2 \xi_x^2} \quad \times \nonumber \\ & \quad \left(\sqrt{(1+2
\xi_F^2)(1 + \tan^2 \zeta + 2\xi_x^2)} -\tan
\zeta\right)^{\frac{1}{2}}\,,
\label{Eq:homcondvarxx} \\
V_{pp} &= \frac{\hbar m \Omega _q}{\sqrt{2}} \sqrt{1+2 \xi_F^2}
\quad \times \nonumber \\ & \quad \left(\sqrt{(1+2
\xi_F^2)(1+\tan^2 \zeta+ 2 \xi_x^2)} - \tan
\zeta\right)^{\frac{1}{2}}\,, \label{Eq:homcondvarpp} \\
V_{xp} &= \frac{\hbar}{2} \left(\sqrt{(1+2 \xi_F^2)(1+\tan^2
\zeta+2 \xi_x^2)}-\tan \zeta\right)\,, \label{Eq:homcondvarxp}
\end{align}
while the purity of the state is given by
\begin{align} \label{Eq:hompurity}
\det\mathbf{V} &= \frac{\hbar^2}{4}\,\left(\left(1+2
\xi_F^2\right) \left(1+2 \xi_x^2\right)+2 \xi_F^2\,
\tan^2\zeta\right) \nonumber \\
&\ge \frac{\hbar^2}{4}\, \left(1+2 \ \xi_F\, \xi_x\right)^2\,.
\end{align}
Here we have defined the two ratios, $\xi_F \equiv
\Omega_F/\Omega_q$ and $\xi_x \equiv \Omega_q/\Omega_x$ with
$\Omega_q = \alpha/\sqrt{\hbar\, m}$. Then we can recover
Eq.\,\eqref{Eq:Uapprox} and with the Appendix\,\ref{Sec:Neff}  we
obtain
\begin{equation} \label{Eq:Neffcond}
\mathcal{N}_{\rm eff} = \frac{S_{\rm cl}(\Omega_{\rm
cl})}{2\,S_{\rm SQL}(\Omega_{\rm cl})}\,.
\end{equation}

In the quantum-noise limit with $\xi_F = \xi_x = 0$ the
conditional state is pure for any measurement frequency $\Omega_q$
and homodyne detection angle $\zeta$. This defines the {\it
conditional ground state of a free mass}
\begin{align}
V_{xx} &= \frac{\hbar }{\sqrt{2}\,m\, \Omega _q}\,,
\label{Eq:quantcondvarxx} \\
V_{pp} &= \frac{\hbar\,m\,\Omega _q}{\sqrt{2}} \,,
\label{Eq:quantcondvarpp} \\
V_{xp} &= \frac{\hbar}{2}\,. \label{Eq:quantcondvarxp}
\end{align}
Here the Wiener filter functions for position and momentum become
equal to simple decaying cosine functions at the measurement
frequency and read
\begin{align}
K_x (t) &= \sqrt{2}\, \Omega _q\, e^{-\frac{\Omega _q\,
t}{\sqrt{2}}}\, \cos \left(\frac{\Omega _q\,t}{\sqrt{2}}\right)\,, \\
K_p (t) &= \sqrt{2}\, m\, \Omega _q^2\, e^{-\frac{\Omega _q\,
t}{\sqrt{2}}}\, \cos \left(\frac{\Omega
_q\,t}{\sqrt{2}}+\frac{\pi}{4}\right)\,.
\end{align}
Then it becomes apparent that the inverse of the measurement
frequency $\Omega_q$ (at which the total noise approaches SQL the
most) is also the time scale at which information regarding
test-mass position and momentum have to be extracted from the
output data.

It is clear from Eq.\,\eqref{Eq:hompurity} that one should measure
the phase quadrature, i.e. $\zeta=0$, in order to minimize the
uncertainty product. That means that in order to obtain a small
uncertainty product, it is not required to remove the quantum
back-action noise from the output, as one would benefit from when
trying to detect gravitational waves~\cite{KLMTV2001}
--- to the contrary, for that purpose it would be even destructive
to do so. This is understandable, since here the aim is to learn
as much as possible about the mirror motion, and the effect of
quantum back action is an important content of the mirror motion.
For example, suppose we chose to measure a back-action-evading
observable using an oscillator with negligible classical noise. In
this case, we would have an output channel that has a minimal
power spectrum, and is hence ideal for measuring any classical
force that acts onto the mirror. Fluctuation in the mirror motion
and the momentum driven by back action, on the other hand, would
almost diverge around the resonant frequency. The output field,
containing absolutely no information about the back action, would
not be able to remove this fluctuation via conditioning. This
would then result in a conditional state equal to the
unconditional state with very large variances.

We learn from
Eqs.\,\eqref{Eq:homcondvarxx}--\eqref{Eq:homcondvarxp} that the
effect of the classical force noise in the conditional variances
is suppressed with a higher measurement frequency (random force
has less time to act and accumulate) while the classical sensing
noise is suppressed with a lower measurement frequency (random
sensing noise has longer time to average out). Moreover, in the
absence of any classical sensing noise the test-mass state becomes
pure with an infinitely strong measurement (in which $\Omega_q$
approaches $+\infty$):  all classical forces acting on the test
mass can be neglected in presence of the strong back-action force
and the test mass reaches the conditional ground state at the
infinite measurement frequency. Vice versa, in the theoretical
absence of classical force noise, the test-mass state becomes pure
in the limit of an infinitely weak measurement (in which
$\Omega_q$ approaches 0): if the test-mass motion is only driven
by the measurement's back action but this motion is then
unfortunately hidden in the measurement output because it is
covered by the classical sensing noise, the best idea would be not
to measure the test mass at all. If both classical noise sources
are present, the uncertainty product is minimized further with an
optimal power which accomplishes a balancing between classical
force and sensing noise, i.e. with $\xi_F = \xi_x$. This produces
an equal sign in the second line of Eq.\,\eqref{Eq:hompurity} and
is true for a measurement frequency of $\Omega_q = \Omega_{\rm
cl}$. This simply means that the quantum noise should touch the
free-mass SQL at the frequency where the classical noise has the
maximal separation to that limit (cf. Fig.\,\ref{Fig:SNDMARKOV}).
The expression of the minimal uncertainty product and the one of
the minimal effective occupation number are then functions of the
classical-noise-SQL-beating factor.

For a mechanical object there is not a fundamental definition of
which state is vacuum, since it is possible to vary the potential
well it lies in. Nevertheless, states with non-zero correlation
among displacement and momentum can always be regarded as
squeezed, and we can often discuss whether one state is more
position squeezed (momentum anti-squeezed) or more position
anti-squeezed (momentum squeezed) compared to another.
\begin{figure}[t]
\centerline{\includegraphics[width=0.7\linewidth,angle=-90]{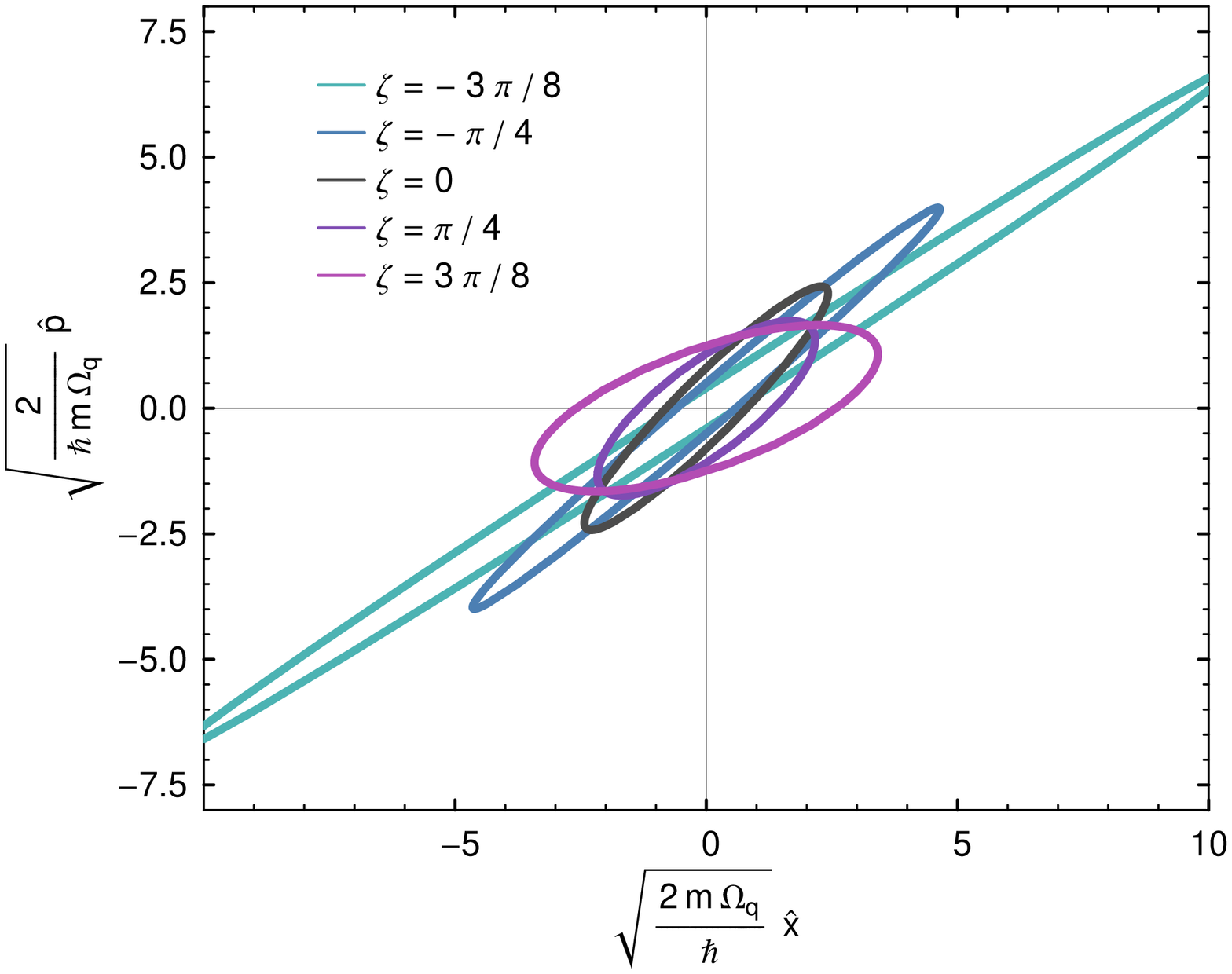}}
\vspace*{0.2cm}
\centerline{\includegraphics[height=0.9\linewidth,angle=-90]{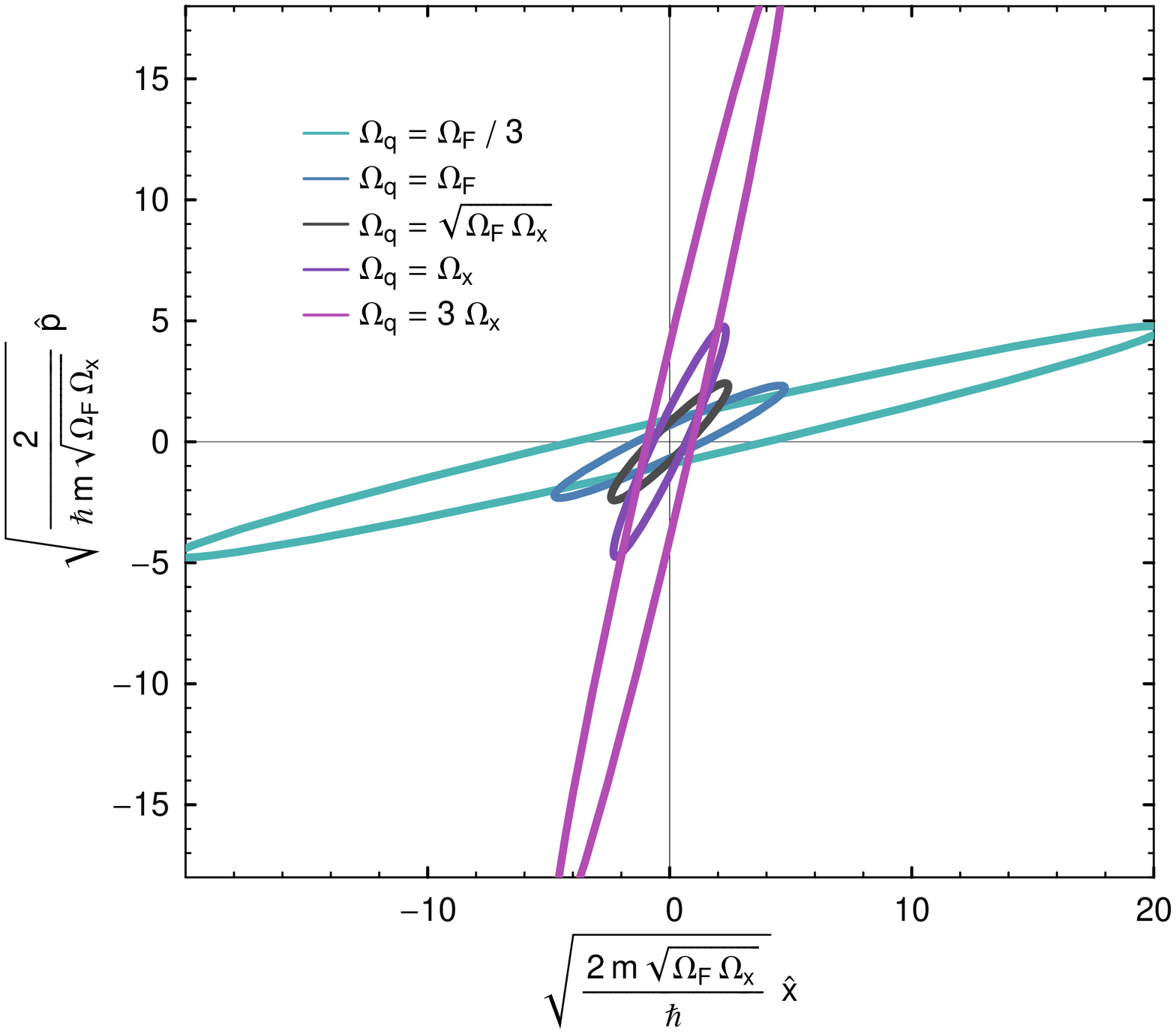}}
\caption{Test-mass squeezing normalized with respect to the
conditional ground state of a free mass for optimal measurement
frequency $\Omega_q = \Omega_{\rm cl} \equiv
\sqrt{\Omega_F\,\Omega_x}$ and different homodyne detection angles
(upper panel) and for different measurement frequencies at phase
quadrature detection (lower panel) including in both cases a
certain classical noise budget: we have chosen $\Omega_x/\Omega_F
= 5$.} \label{Fig:sqhom}
\end{figure}
We have illustrated the squeezing situation, by plotting noise
ellipses obtained with various homodyne detections (see upper
panel of Fig.\,\ref{Fig:sqhom}). As the homodyne detection angle
$\zeta$ vary from optimal value $0$ to $-\pi/2$, the semimajor
axis of the noise ellipse, i.e. the anti-squeezed quadrature,
becomes rotated into the direction of the position. At the same
time the ellipse becomes more stretched, i.e. the squeezed
quadrature is more squeezed, while the anti-squeezed quadrature is
more anti-squeezed. As $\zeta$ vary from $0$ to $\pi/2$, the
semimajor axis of the noise ellipse also rotates into the
direction of the position but the ellipse becomes rather bulged.
Furthermore, we show how the noise ellipse changes, as the
measurement frequency is not chosen to be equal to the geometrical
mean of the two classical noise frequencies, i.e. $\Omega_q \neq
\Omega_{\rm cl}$. For phase quadrature detection as in the lower
panel of Fig.\,\ref{Fig:sqhom}, a slow measurement, i.e. with a
low measurement frequency $\Omega_q < \Omega_{\rm cl}$, generates
a position anti-squeezed (and momentum squeezed) conditional
state; while a fast measurement with $\Omega_q > \Omega_{\rm cl}$
generates a position squeezed (and momentum anti-squeezed)
conditional state. In addition, the deviation from the optimal
measurement frequency, i.e. $\Omega_q \neq \Omega_{\rm cl}$,
always produces a less pure states.

\subsection{Very low finesse cavity and free mass scenario
with squeezed vacuum input}

Up to now we have only treated in-going coherent vacuum states but
one could also think about squeezed vacuum states coupling to the
mirror~\cite{KLMTV2001}. This corresponds to inserting squeezed
states into an interferometer's dark port. By doing so the quantum
limited sensitivity of an interferometer can be
enhanced~\cite{Cav1980}. The free-mass conditional state is given
by
\begin{align}
V_{xx} &= \frac{\hbar}{\sqrt{2} m \Omega_q}\sqrt{\lambda_-^2+2
\xi_x^2} \quad \times \nonumber \\
& \quad \left(\sqrt{(\lambda_+^2+2 \xi_F^2) (\lambda_-^2+2
\xi_x^2)} -\sin 2\varphi_{\rm op}\,\sinh 2r_{\rm
op}\right)^{\frac{1}{2}}\,,
\label{Eq:sqcondvarxx} \\
V_{pp} &= \frac{\hbar m \Omega_q}{\sqrt{2}}\sqrt{\lambda_+^2+ 2
\xi_F^2} \quad \times \nonumber \\ & \quad
\left(\sqrt{(\lambda_+^2+2 \xi_F^2) (\lambda_-^2+2 \xi_x^2)} -\sin
2\varphi_{\rm op}\,
\sinh 2r_{\rm op}\right)^{\frac{1}{2}}\,, \label{Eq:sqcondvarpp} \\
V_{xp}&= \frac{\hbar}{2}\left(\sqrt{(\lambda_+^2+2 \xi_F^2)
(\lambda_-^2+2 \xi_x^2)}-\sin 2\varphi_{\rm op}\,\sinh 2r_{\rm
op}\right)\,. \label{Eq:sqcondvarxp}
\end{align}
Here we have defined $\lambda_{\pm}^2 = \cosh 2 r_{\rm op}\,\pm
\cos 2 \varphi_{\rm op}\, \sinh 2 r_{\rm op}$, where $(20/\ln
10)\,r_{\rm op}>0$ gives the optical squeezing strength in dB at a
squeezing angle of $\varphi_{\rm op}$. Then the purity of the
conditional state can be inferred from
\begin{align} \label{Eq:sqpurity}
\det {\bf V} &= \frac{\hbar^2}{4}\, \left((\lambda_+^2+2 \xi_F^2)
(\lambda_-^2+2 \xi_x^2)-\sin^2 2\varphi_{\rm
op}\,\sinh^2 2r_{\rm op}\right) \nonumber \\
&\geq \frac{\hbar^2}{4}\, \Big(1+4\, \xi _F^2\, \xi _x^2 + 2\,(\xi
_F^2+\xi _x^2)\, \cosh 2 r_{\rm op} \nonumber \\ & \quad -2\,|\xi
_F^2-\xi _x^2|\,\sinh 2r_{\rm op}\Big) \ge \frac{\hbar^2}{4}
(1+2\, \xi_F\, \xi_x)^2\,.
\end{align}
The equality of the first inequality sign in
Eq.\,\eqref{Eq:sqpurity} is achieved at $\varphi_{\rm op} = 0$ for
$\xi_F^2 > \xi_x^2$ and at $\varphi_{\rm op} = \pi/2$ for $\xi_F^2
< \xi_x^2$, i.e. by squeezing either the phase or the amplitude
quadrature, respectively. Note that in Eq.\,\eqref{Eq:sqpurity}
for any $\Omega_q$, the same minimum as in
Eq.\,\eqref{Eq:hompurity} is reached if $r_{\rm op} = {\rm
arctanh}(|\xi _F^2-\xi _x^2|/(\xi _F^2+\xi _x^2))/2$ --- even when
having $\xi _F \not= \xi _x$. Therefore, even with input
squeezing, the conditional state cannot become more pure than with
coherent input, but the demands on the required measurement
frequency --- and with this the constraints on the optical power,
which is needed in order to obtain a certain uncertainty product,
can be relaxed. In real experiments the optical power is of course
always limited and squeezed input becomes an very important tool.

It has turned out that the conditional variances are in principle
even analytically equivalent in the following two cases: {\it (i)}
input-squeezing at a flexible but frequency-independent angle or
{\it (ii)} flexible amount of available optical power and a
flexible but frequency-independent homodyne detection angle. This
can easily be seen by replacing the homodyne detection angle $\tan
\zeta \rightarrow \sin 2\varphi_{\rm op} \sinh 2r_{\rm op}$ and
the measurement frequency $\Omega_q \rightarrow \lambda_+
\Omega_q$ in
Eqs.\,\eqref{Eq:homcondvarxx}--\eqref{Eq:homcondvarxp}. Then we
simply end up with
Eqs.\,\eqref{Eq:sqcondvarxx}--\eqref{Eq:sqcondvarxp}. Here we can
directly see that using input squeezing allows to change the
parameters within $\Omega_q$ such as the optical power, the laser
frequency and the mirror mass but by modifying the input squeezing
parameter $\lambda_+$ we can at the same time maintain the
measurement frequency.

Even though a homodyne detection different from the phase
quadrature and input squeezing do both not help with increasing
the purity of the conditional state they increase the squeezing of
the conditional test-mass state. Furthermore, with a certain
homodyne detection angle or with a certain input squeezing it is
possible to minimize the position and momentum correlation in the
conditional state.

\section{Macroscopic entanglement}
\label{Sec:MacroEnt}

The concept of entangled states is one of the most important
phenomenons when entering the quantum regime. In
Sec.\,\ref{SubSec:FinCav} we will see that the entanglement
between the cavity mode and the mirror motion is in fact
responsible for a degeneration of the purity of the mirror's
quantum state. In this section we will explain how we can prepare
quantum entanglement in position and momentum between the centers
of mass of the two end mirrors in the north $\hat x^{\rm n}$ and
the east $\hat x^{\rm e}$ arm of a simple, but power-recycled
Michelson interferometer using the conditional states as derived
in the previous section. The end mirrors are suspended as
pendulums but with a very low eigenfrequency. Therefore, such an
experiment would be in direct analogy to the
Einstein-Podolsky-Rosen gedanken experiment~\cite{EPR1935}. Here
we will basically extend the discussion carried out in
Ref.~\cite{MRSDC2007}.

Recall that the common $(\hat x^{\rm e} + \hat x^{\rm n})$ and the
differential $(\hat x^{\rm e} - \hat x^{\rm n})$ mode of motion
between the two end mirrors are independent and can each be sensed
by a homodyne detection at the bright and the dark port,
respectively, as suggested in Ref.~\cite{MRSDC2007}. As already
mentioned before, if using the reduced mass of the mirrors, the
conditional variances as derived in Sec.\,\ref{Sec:Marko} hold for
the differential mode observed at the dark port of the
interferometer. But in order to describe the common mode they have
to be slightly modified since the power-recycling cavity --- here
with high bandwidth and therefore an adiabatically eliminated
cavity mode --- with transmissivity $\tau$ enhances the
measurement strength $\alpha^{\rm c} = 2/\tau\, \alpha > \alpha$
and is therefore different to the one associated with the
differential mode $\alpha^{\rm d} = \alpha$. Furthermore, the
common mode will suffer additionally to the classical force noise
and the classical sensing noise --- note that we suppose that
these two classical noise sources are equally distributed into
common and differential mode --- from laser noise, since the
in-going modulation fields at the bright port are usually not in
vacuum states. We have to make the following additional
replacements in
Eqs.\,(\ref{Eq:homcondvarxx})--(\ref{Eq:homcondvarxp}) in order to
obtain the conditional variances of the common mode: $(1+2\xi_F^2)
\rightarrow (S_{a_1a_1}+2\xi_F^2)$ and $(1+\tan^2 \zeta +
2\xi_x^2) \rightarrow (S_{a_2a_2}+ S_{a_1a_1} \tan^2 \zeta +
2\xi_x^2)$ as well as replacing the detached $(- \tan \zeta)$ in
each variance by $(- S_{a_1a_1} \tan \zeta)$. Here
$S_{a_{1}a_{1}},\,S_{a_{2}a_{2}} \geq 1$ are the (frequency
independent) spectra of the technical laser noise in amplitude and
phase, respectively.

Then we can assemble the conditional state of the entire system:
the combined covariance among $(x^{\rm e}, p^{\rm e}, x^{\rm n},
p^{\rm n})$ simply reads
\begin{equation}
{\bf V}_{\rm total} = \left(\begin{array}{cc} {\bf V}_{\rm ee} &
{\bf V}_{\rm en} \\ {\bf V}_{\rm ne} & {\bf V}_{\rm nn}
\end{array} \right)
\end{equation}
with
\begin{eqnarray}
{\bf V}_{\rm nn}={\bf V}_{\rm ee}&=&  \left(\begin{array}{cc}
(V_{xx}^{\rm c} + V_{xx}^{\rm d})/4
& (V_{xp}^{\rm c} + V_{xp}^{\rm d})/2 \\
(V_{xp}^{\rm c} + V_{xp}^{\rm d})/2 & V_{pp}^{\rm c} + V_{pp}^{\rm
d} \\ \end{array} \right)\,, \nonumber \\
{\bf V}_{\rm en}={\bf V}_{\rm ne}&=&  \left(\begin{array}{cc}
(V_{xx}^{\rm c} - V_{xx}^{\rm d})/4 & (V_{xp}^{\rm c} -
V_{xp}^{\rm d})/2
\\ (V_{xp}^{\rm c} - V_{xp}^{\rm d})/2 & V_{pp}^{\rm
c} - V_{pp}^{\rm d}\\ \end{array} \right)\,. \nonumber
\end{eqnarray}
This combined covariance matrix is very similar to the covariance
matrix for the amplitude and the phase quadrature of two output
light beams which have been created by overlapping two continuous
Gaussian light beams on a beam splitter. Note that overlapping two
light beams which are differently squeezed in amplitude and phase
quadrature on a beam splitter is a very common way of how
continuous variable entanglement is created in
optics~\cite{FSBFKP1998,BSL2003}. In the mirror case the common
and the differential mode are mathematically overlapped to give
the motion of each individual end mirror.

In the following we will use the {\it logarithmic negativity},
which was introduced for an arbitrary bipartite system in
Ref.~\cite{ViWe2002}, as a quantitative measure of the
entanglement. For our state it reads
\begin{equation} \label{logneg}
E_{\mathcal N} = \max[0, - \log_2 2 \sigma^-/\hbar],
\end{equation}
where we have $\sigma^- = \sqrt{(\Sigma - \sqrt{\Sigma^2 - 4 \det
V})/2}$ and $\Sigma = \det V_{\rm nn} + \det V_{\rm ee} - 2 \det
V_{\rm ne} = V_{xx}^{\rm c}  V_{pp}^{\rm d} + V_{pp}^{\rm c}
V_{xx}^{\rm d} - 2 V_{xp}^{\rm c}  V_{xp}^{\rm d}$. The higher the
value of $E_{\mathcal N}$ the stronger the entanglement.

Recall that there exists a frequency band with sub-SQL classical
noise if $\Omega_x/\Omega_F > 2$. However, the existence of
entanglement sets a slightly higher threshold value for this
frequency ratio depending on the strength of laser noise as it is
shown in Ref.~\cite{MRSDC2007}. We know from Sec.\,\ref{Sec:Marko}
that the uncertainty product of each individual mode --- common
and differential --- is minimal for a phase quadrature detection
$\zeta^{\rm c,d}=0$ but it has turned out that this is not the
optimal choice for the preparation of entanglement. If the
homodyne detection angle approaches $-\pi/2$, each mode can become
more squeezed depending on the measurement frequency.

The entanglement between the two mirrors --- created by
overlapping two modes --- increases with the squeezing of the
individual modes and with the angle separating the squeezed
quadrature of the two modes.
\begin{figure}[t]
\centerline{\includegraphics[height=\linewidth,angle=-90]{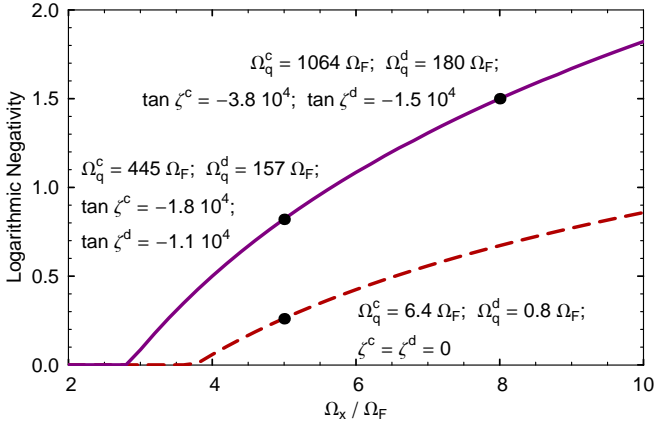}}
\caption{Logarithmic negativity versus $\Omega_x/\Omega_F$
maximized with respect to $\Omega_q^{\rm c}$ and $\Omega_q^{\rm
d}$ using phase quadrature detection (dashed red line) as well as
additionally maximized with respect to $\zeta^{\rm c}$ and
$\zeta^{\rm d}$ (solid purple line). In both cases no laser noise
is assumed, i.e. with $S_{a_1a_1}=S_{a_2a_2}=1$. At some positions
optimal parameter values for $\Omega_q^{\rm c}$, $\Omega_q^{\rm
d}$, $\zeta^{\rm c}$ and $\zeta^{\rm d}$ are given in the plot.}
\label{Fig:lnmax}
\end{figure}
Then it is obvious that one should not observe common and
differential mode via phase quadrature detection, but that there
is a certain value, for each the common and the differential mode,
of $\Omega_\alpha^{\rm c,d}$ and $-\pi/2 < \zeta^{\rm c,d} < 0$
which is optimal for the entanglement and maximizes the
logarithmic negativity (cf. the solid line in
Fig.\,\ref{Fig:lnmax}). These optimal parameters depend of course
on the classical noise, but are usually characterized by a high
measurement frequency and a detection close to $-\pi/2$ for both
modes. That means that the states are totally driven by radiation
pressure which is in turn monitored by reading out close to the
amplitude quadrature.

Furthermore, we have found that the laser noise --- entering at
the bright port and in a totally balanced interferometer only
affecting the common mode --- can theoretically be almost
suppressed with the optimally high measurement frequency and the
optimal homodyne detection angle. Therefore, the resulting maximal
entanglement --- represented by the solid line in
Fig.\,\ref{Fig:lnmax} --- is independent of the laser noise. But
the parameters --- such as optical power and the fine-tuning of
the homodyne detection angle, which are required in order to reach
the maximal entanglement, are different for different strength of
laser noise and are far away from any realistic situation (cf. the
dots in Fig.\,\ref{Fig:lnmax}).

\section{Interferometers with non-Markovian noise}
\label{Sec:NonMarko}

\subsection{Cavity with finite bandwidth}
\label{SubSec:FinCav}

In realistic experimental situations, the noise sources are usually not Markovian. 
In this section we want to generally study quantum state preparation 
in the background of non-Markovian noise sources. We start with allowing 
the quantum noise spectral density to become frequency dependent.

If we consider a cavity of length $L$ with a finite half cavity
bandwidth $\gamma$ and a movable end mirror, the quantum noise is
indeed non-Markovian and the Heisenberg equations of motion in frequency
domain modify to
\begin{eqnarray}
\hat y &=& \sin \zeta\, \frac{\gamma + {\rm i} \Omega}{\gamma -
{\rm i} \Omega}\, \hat a_1 + \cos \zeta \times \nonumber \\
& & \qquad \left[\frac{\gamma + {\rm i} \Omega}{\gamma - {\rm i}
\Omega}\,\hat a_2 + \frac{\sqrt{2c\gamma/L}}{\gamma - {\rm i}
\Omega}\,\frac{\alpha}{\hbar} (\hat x (\Omega) + \hat
\xi_x)\right]\,, \label{Eq:eomcy}\\
\hat x &=& - \frac{1}{m (\Omega^2 + i\gamma_m\Omega-\omega_m^2)}
\left[\frac{\sqrt{2c\gamma/L}}{\gamma - {\rm i} \Omega}\alpha\,
\hat a_{1} + \hat \xi_F\right]\,, \label{Eq:eomcx}
\end{eqnarray}
where we have approximated exponential functions by rational
functions of $\Omega$. Now the measurement frequency becomes
\begin{equation}\label{eqn:mfreqcav}
\Omega_q^{\rm cav} = \sqrt{\frac{2\,c}{m\,\hbar\,L\,\gamma}}\,
\alpha \,,
\end{equation}
which corresponds to the frequency where the quantum noise
spectral density of the associated adiabatically eliminated system
touches the SQL. If the phase quadrature of the outgoing light
($\zeta = 0$) is detected, the eight zeros of its spectral density
$S_{yy}$ are given by $\pm a_1 \pm {\rm i} b_1$ and $\pm a_2 \pm
{\rm i} b_2$  with (for simplicity for a free mass, i.e.
$\gamma_m=\omega_m= 0$):
\begin{align}
a_{1,2} =& \frac{1}{2}
\sqrt{\sqrt{r^2\mp\sqrt{2}r}\pm\frac{r}{\sqrt{2}}-1}\,, \\
b_{1,2} =& \frac{1}{2}
\sqrt{\sqrt{r^2\mp\sqrt{2}r}\mp\frac{r}{\sqrt{2}}+1}\,,
\end{align}
where $r=\sqrt{\sqrt{\left(2\Omega_q^{\rm
cav}/\gamma\right)^4+1}+1}$. The zeros are required for the
spectral decomposition introduced in Sec.\,\ref{Sec:Wiener}. After
straightforward algebraic manipulations, one  arrives at the
following conditional second-order moments:
\begin{align}
V_{xx} =& \frac{\hbar \gamma}{6 m (\Omega_q^{\rm cav})^2} (c_1^3\!
+\! 3 c_1^2\! +\! 3 c_1\! +\!
c_3)\,, \label{eqn:vxxcav} \\
V_{pp} =& \frac{\hbar m \gamma^3}{120 (\Omega_q^{\rm cav})^2}
(3 c_1^5\! +\! 15 c_1^4\! +\! 20 c_1^3\! +\! 60 c_3\! +\! 60c_5)\,, \label{eqn:vppcav}\\
V_{xp} =& \frac{\hbar \gamma^2}{16 (\Omega_q^{\rm cav})^2} c_1^2
(c_1\! +\! 2)^2\label{eqn:vxpcav}\,,
\end{align}
where the coefficients are given by
\begin{equation}
c_n = \frac{2}{n}\Im\left[(a_1+{\rm i}b_1)^n+(a_2+{\rm
i}b_2)^n-{\rm i}^n\right]\,.
\end{equation}

The conditional variances given in
\begin{figure}
\includegraphics[height=\linewidth,angle=-90]{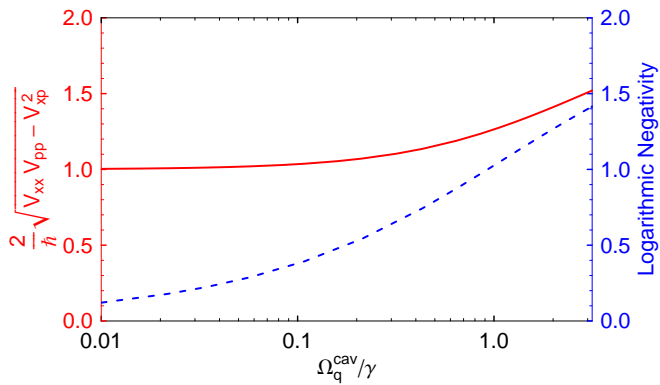}
\caption{Test-mass uncertainty product $U$ (solid line) compared
to the entanglement between test mass and cavity mode (dashed
line) both versus the dimensionless ratio $\Omega_q^{\rm
cav}/\gamma$ through which both quantities are totally described.
No classical noise is present. Free mass limit is used, i.e.
$\gamma_m=\omega_m=0$, and the phase quadrature $\zeta=0$ is
detected. Entanglement is quantified by the logarithmic
negativity.} \label{Fig:fbq}
\end{figure}
Eqs.\,\eqref{eqn:vxxcav}-\eqref{eqn:vxpcav}, although slightly
complicated, are still analytic; we can still draw some important
conclusions from these second-order moments. By expanding the
quantity $U$ from Eq.\,\eqref{Eq:Uquant} in terms of
$\Omega_q^{\rm cav}/\gamma$, we obtain
\begin{equation}
U =1\! +\! \frac{1}{2\sqrt{2}}\frac{\Omega_q^{\rm cav}}{\gamma}\!
+\! \mathcal{O}\left((\Omega_q^{\rm cav}/\gamma)^2\right)\,,
\end{equation}
which reveals that, even in the quantum noise limited case, the
conditional state of the test mass cannot be pure as long as
$\Omega_q^{\rm cav}/\gamma > 0$. This is in contrast to the
Markovian limit ($\gamma \rightarrow \infty$) discussed in
Sec.\,\ref{Sec:Marko}, where the conditional state is always pure
in the absence of classical noise [cf. Eq.\,(\ref{Eq:hompurity})].
Fig.\,\ref{Fig:fbq} shows the purity of the test mass versus the
dimensionless ratio $\Omega_q^{\rm cav}/\gamma$.

In the case of a finite cavity bandwidth, the light is stored
inside the cavity for some time. The information carried by the
light concerning the test massís state cannot leave the cavity
instantaneously and hence is not accessible for the conditioning
process. Consequently, the intra-cavity mode needs to be taken
into account for a complete characterization of the system. The
residual second-order moments required for completing the
corresponding $(4\times 4)$ conditional covariance matrix can be
obtained in the same way as
Eqs.\,\eqref{eqn:vxxcav}-\eqref{eqn:vxpcav}. It turns out that the
composite system is indeed a pure one even though each individual
system resides in a mixed state. This is a clear evidence of
entanglement between the conditional states of the test mass and
the cavity mode. We have also plotted the logarithmic negativity
in Fig.\,\ref{Fig:fbq}. The test-mass state's purity decreases and
test-mass-light entanglement increases with smaller bandwidth and
with higher measurement frequency $\Omega_q^{\rm cav}$. Note that
the uncertainty product as well as the logarithmic negativity do
not diverge. But Fig.\,\ref{Fig:fbq} indicates that, as long as
$\Omega_q^{\rm cav} \ll \gamma$, we can neglect this effect and
adiabatically eliminate the cavity mode as performed for the
power-recycling cavity in Sec.\,\ref{Sec:MacroEnt}.

\begin{figure}
\includegraphics[height=0.8\linewidth,angle=-90]{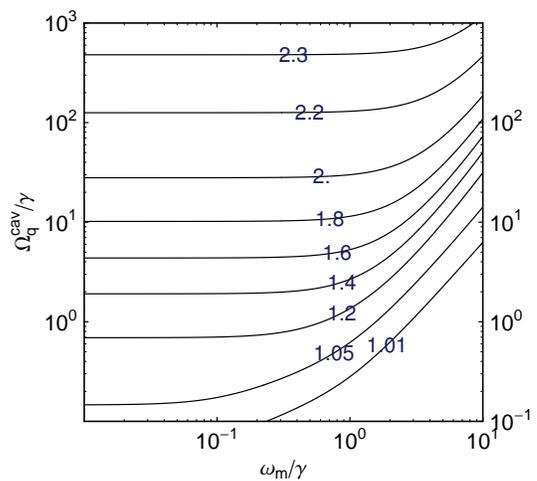}
\caption{Contour plot of the normalized test-mass uncertainty
product given by $U = 2/\hbar \sqrt{V_{xx} V_{pp} - V_{xp}^2}$
versus $\Omega_q^{\rm cav}/\gamma$ and $\omega_m/\gamma$ for
quantum noise only. Again phase quadrature ($\zeta=0$) is
detected.} \label{Fig:fbqm}
\end{figure}

Fig.\,\ref{Fig:fbqm} further shows, that the purity increases with
higher mechanical eigenfrequency $\omega_m$, depending on the
measurement frequency.
Let us consult the following hand-waving argument: with increasing
$\omega_m$ the mechanical oscillator and the optical oscillator,
which would resonate at modulation-frequency zero, become more
separated in the frequency space and therefore their entanglement
decreases. And with decreasing entanglement the test-mass state
becomes more pure.
\begin{figure}
\centerline{\includegraphics[width=\linewidth,angle=0]{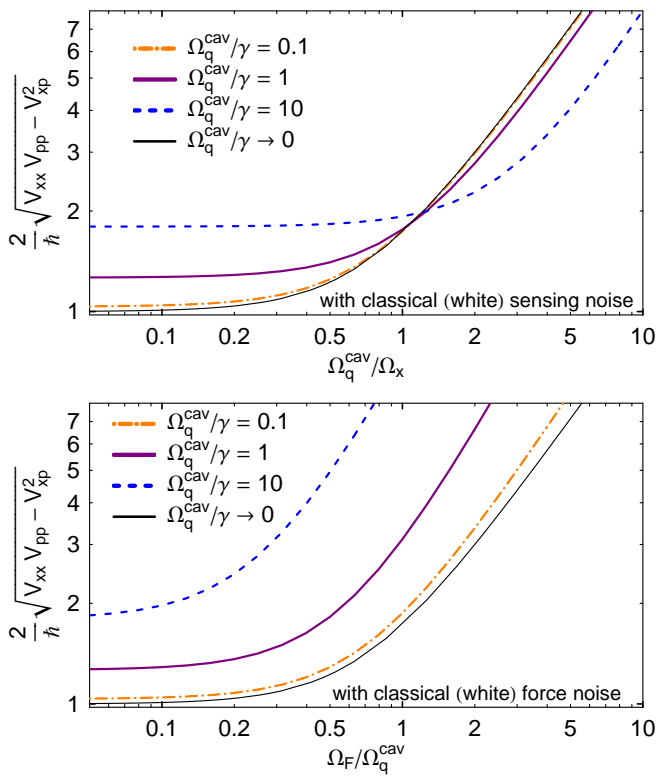}}
\caption{Test-mass uncertainty product $U$ versus classical
sensing noise and without force noise (upper panel) as well as
versus classical force noise without classical sensing noise
(lower panel), both for different examples of ratios between
measurement frequency and bandwidth. Free mass limit is used, i.e.
$\omega_m=0$, as well as phase quadrature detection $\zeta=0$.}
\label{Fig:fbcl}
\end{figure}
But the regime with such high mechanical resonance frequencies is
usually not available in actual GW detectors.

When including the two classical noise sources from our simple
model the conditional state becomes more and more mixed with an
increasing classical noise level as shown in Fig.\,\ref{Fig:fbcl}.
Note that this is of course not an effect of entanglement, the
classical noise rather destroys the quantum entanglement between
the test mass and the cavity mode. The purity also depends on the
ratio between the measurement frequency and the optical bandwidth.
Recall that the classical force noise increases with higher
$\Omega_F$ while the classical sensing noise increases with lower
$\Omega_x$. If we only take sensing noise into account we know for
sure that the motion of the test mass is solely driven by quantum
back-action noise. A high sensing noise level randomizes the
measurement record and hence the delimitated accessibility of the
intra-cavity field due to the finite bandwidth is insignificant.
Consequently all curves roughly coincide for $\Omega_q^{\rm
cav}\gtrsim\Omega_x$ in the upper panel of Fig.\,\ref{Fig:fbcl}.

Note that in the planned Advanced LIGO detector~\cite{advLIGO},
the measurement frequency $\Omega_q^{\rm cav}$ is planned to
roughly coincide with the half cavity bandwidth at $\gamma/(2\pi)
\sim 100$\,Hz --- corresponding to the solid lines in
Fig.\,\ref{Fig:fbcl}. Furthermore, we expect the suspension
thermal noise to have a $\Omega_F/(2\pi) \sim 30-40$\,Hz --- that
would be less than $\Omega_q^{\rm cav}/2$ --- but the coating
thermal noise may provide a $\Omega_x$ that only coincide with
$\Omega_F$ or is just marginally higher. We can infer form
Fig.\,\ref{Fig:fbcl} that the quantum state of the
interferometer's differential mode is mainly constrained by
classical sensing noise which entails a lower bound of $U \gtrsim
5$ and gives an effective occupation number of $\mathcal{N}_{\rm
eff} \gtrsim 2$. For a more detailed discussion see
Sec.\,\ref{Sec:AdvLIGO}.

\subsection{Detuned cavity}

A cavity which is detuned by $\Delta$ from the carrier's frequency
makes the power inside the cavity also dependent on the motion of
the test-mass mirrors.
\begin{figure}[h]
\includegraphics[height=0.8\linewidth,angle=-90]{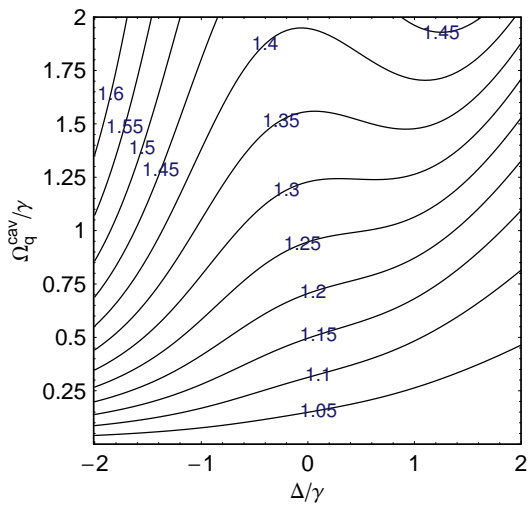}
\caption{Contour plot of the normalized test-mass uncertainty
product given by $U = 2/\hbar \sqrt{V_{xx} V_{pp} - V_{xp}^2}$
versus the ratio between detuning $\Delta$ and bandwidth and
$\Omega_q^{\rm cav}/\gamma$ for quantum noise only. Free mass
limit is used, i.e. $\omega_m=0$, as well as phase quadrature
detection $\zeta=0$.} \label{Fig:fbqd}
\end{figure}
This creates an {\it optical spring}~\cite{BuCh2001} or an optical
anti-spring both shifting the (free) mechanical and the (free)
optical resonance frequency in the complex plane. In future GW
detectors such as Advanced LIGO~\cite{advLIGO}, the optical spring
effect will be used to up-shift the real part of the mechanical
resonance frequency into the detection band. Recall that the
optical spring as well as the optical anti-spring usually
introduces instability to the system which has to be cured with an
appropriate linear feedback control~\cite{BuCh2002}. But it is
straightforward to show that the conditional covariance matrix
does not change under any ideal, linear feedback control.

The Heisenberg equations of motion for such a system can be found
in many previous works, see e.g. Eqs.\,(39)--(41) of
Ref.\,\cite{BuCh2003}. Unfortunately, analytic expressions for the
conditional covariance matrix are cumbersome, and we only report
numerical results. We also restrict ourselves to quantum noise,
and reading out the phase quadrature, i.e. $\zeta=0$ (cf.
Eq.\,(3.2) of Ref.\,\cite{BuCh2002}). Note that the measurement
frequency $\Omega_q^{\rm cav}$ is defined in the same way as for
the tuned finite bandwidth configuration (cf.
Eq.\,\eqref{eqn:mfreqcav}).

Fig.\,\ref{Fig:fbqd} shows that detuning a cavity from the carrier
frequency properly, can increase the purity which comes from the
fact that the quantum entanglement between test mass and cavity
mode is decreased. In the regime of a blue detuned cavity ($\Delta
> 0$) --- producing an optical spring --- and for $\Omega_q^{\rm
cav} < \gamma$, Fig.\,\ref{Fig:fbqd} simply agrees with
Fig.\,\ref{Fig:fbqm}. Here at fixed measurement frequency
$\Omega_q^{\rm cav} < \gamma$ a higher detuning $\Delta$ gives a
less shifted mechanical resonance, so-called {\it optomechanical}
resonance, and at the same time it corresponds to a higher optical
resonance. Therefore, again the two oscillators are more separated
in the frequency space and their entanglement decreases.
Interestingly, for higher $\Omega_q^{\rm cav}$ the test-mass state
could locally appear more pure in the red detuned cavity regime,
i.e. at a certain $\Delta < 0$, which produces an optical
anti-spring. Note that the uncertainty product diverges for an
infinitely red detuned cavity ($\Delta \rightarrow -\infty$). For
these facts we unfortunately have not found any intuitive
explanation. This will be a subject of further investigation.

\subsection{Non-Markovian classical noise}
\label{SubSec:nonmarkonoise}

In the following we consider a more realistic example
configuration involving multiple colored classical noise sources.
The classical noise contributions are highly non-Markovian and
they tend to rise fast in the low frequency regime, which is
ignored by a simple Markovian noise model.
\begin{figure}[t]
\centerline{\includegraphics[height=\linewidth,angle=-90]{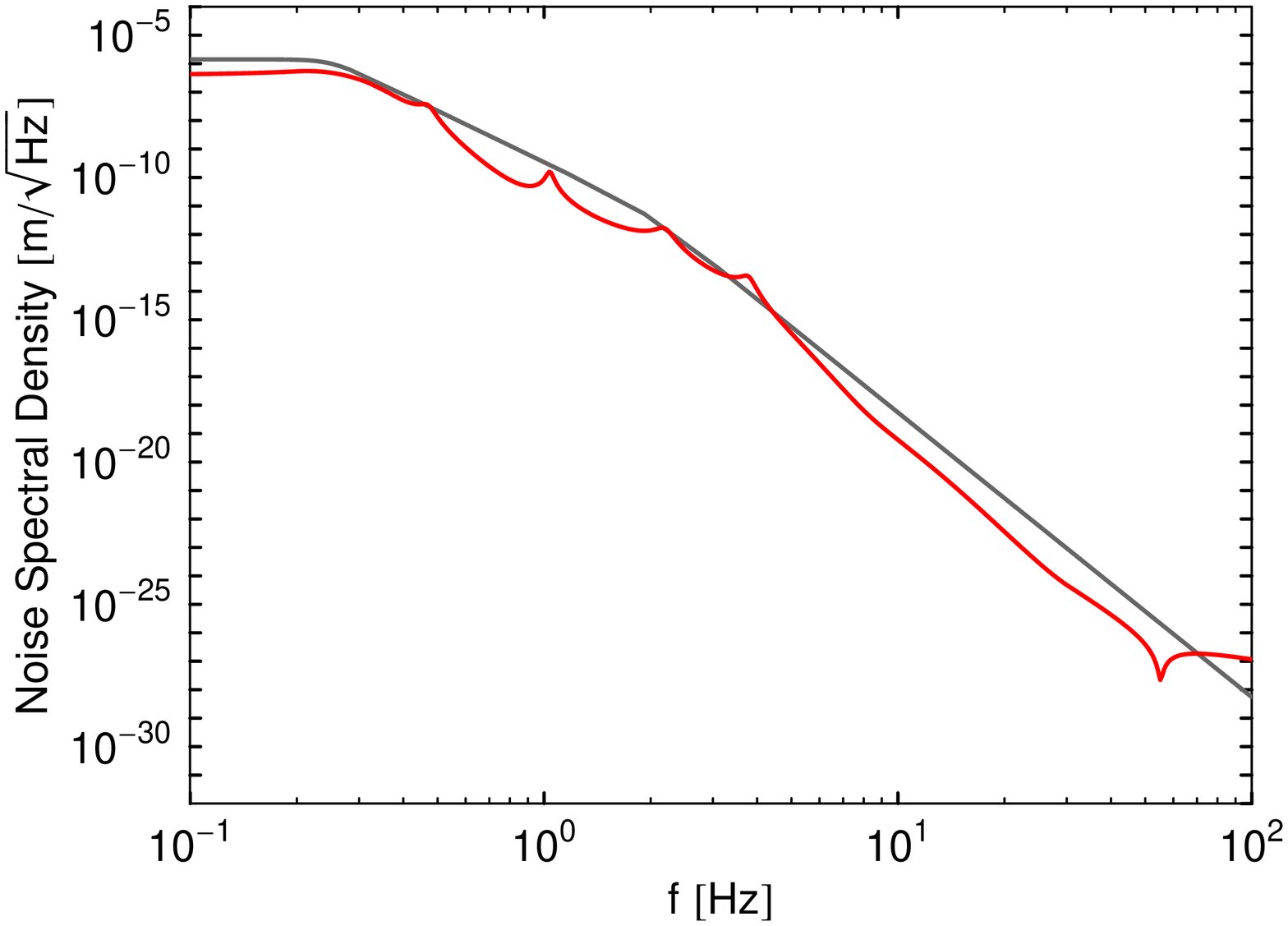}}
\vspace*{0.03\linewidth}
\centerline{\includegraphics[height=\linewidth,angle=-90]{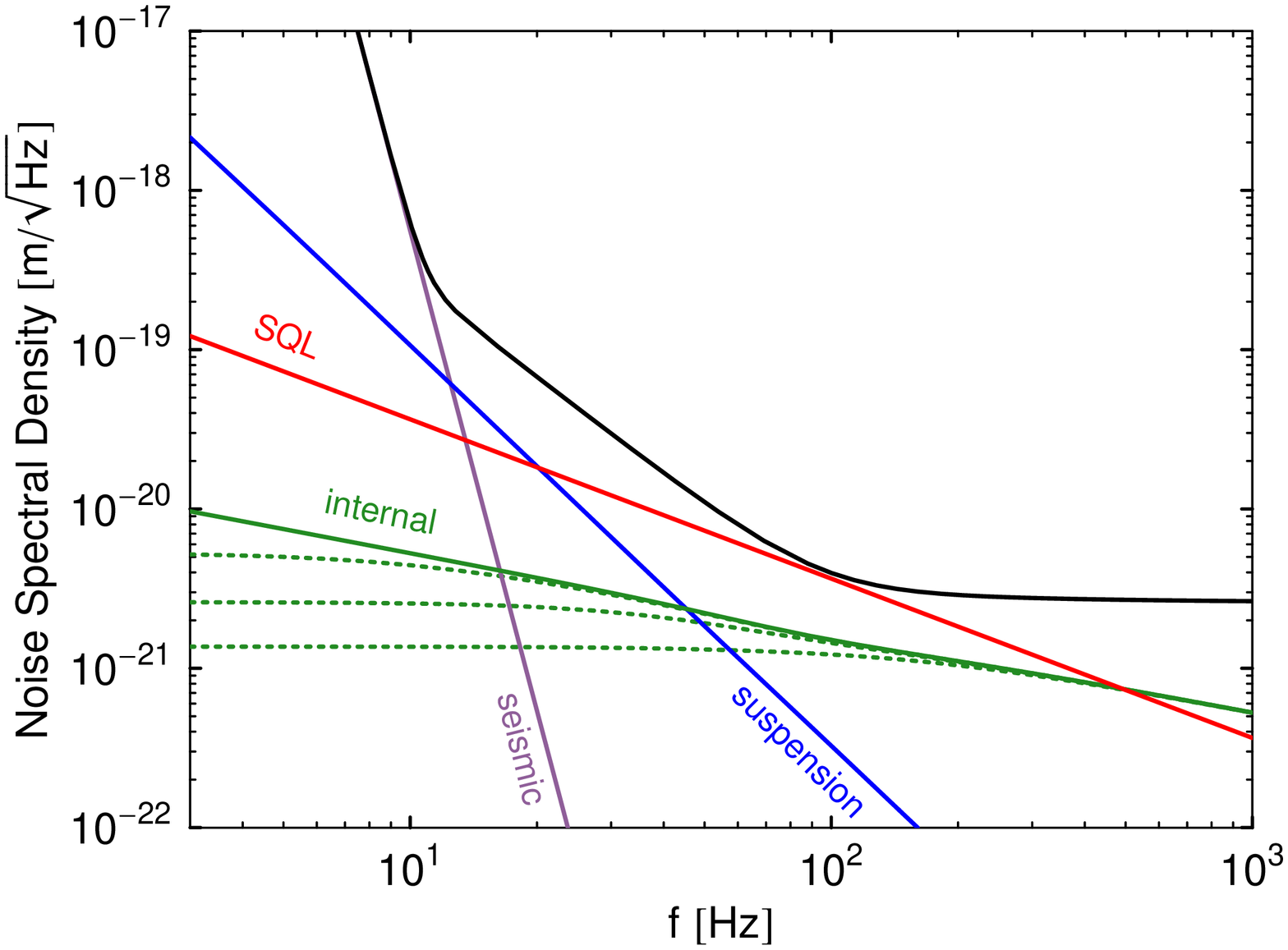}}
\caption{\emph{Upper panel}: seismic noise pre-estimated by the
simulation tool \emph{Bench}~\cite{bench} (red) and a fit by
rational function (gray). \emph{Lower panel}: straw man (colored)
classical noise budget of an advanced interferometric
gravitational wave detector: Seismic (violet), suspension thermal
(blue, follows $\sim 1/f^{5/2}$) and internal thermal (green,
follows $\sim 1/f^{1/2}$) noise spectra are shown as well as the
total noise (black) --- including Markovian quantum noise with
phase quadrature readout. For the thermal noise sources we
employed a Pad\'{e} approximation that generated the $1/f$ spectra
between $0.1 \ {\rm Hz}$ and $2 \ {\rm kHz}$.}\label{fig:cuttoff1}
\end{figure}
First we restrict ourselves to an idealized noise budget of an
advanced interferometric gravitational wave detector, shown in the
lower panel of Fig.\,\ref{fig:cuttoff1}. Here only the dominating
force and sensing noise sources are considered. Additionally we
assume that the gravity gradient noise can be suppressed
completely through monitoring the ground's motion. Especially
seismic noise dominates the entire spectrum below $10 \ {\rm Hz}$.
In order to apply the numerical Wiener filter procedure (cf.
Sec.\,\ref{Sec:Wiener}), all classical noise spectra need to be
approximated by rational functions of $\Omega^2$. This is
illustrated by Fig.\,\ref{fig:cuttoff1} (upper panel) where the
seismic noise spectral density, pre-estimated by the simulation
tool \emph{Bench}~\cite{bench}, is approximated accordingly. The
seismic noise spectrum is constant below $0.25 \ {\rm Hz}$, then
it drops as $\sim 1/f^6$ between $0.25 \ {\rm Hz}$ and $2 \ {\rm
Hz}$ and finally it drops as $\sim 1/f^{10}$ above $2 \ {\rm Hz}$.
The suspension thermal noise constitutes a second force noise
contribution which drops as $\sim 1/f^{5/2}$ above the pendulum
eigenfrequency (at $1\ {\rm Hz}$) and it intersects the SQL at $20
\ {\rm Hz}$. Such a frequency dependance presumes structural
damping. Above $3 \ {\rm Hz}$ the internal thermal noise follows
$\sim 1/f^{1/2}$ and it intersects the SQL at $500 \ {\rm Hz}$. We
have employed the Pad\'{e} expansion in order to simulate the
behavior of the spectral densities.

\begin{figure}[t]
\includegraphics[height=\linewidth,angle=-90]{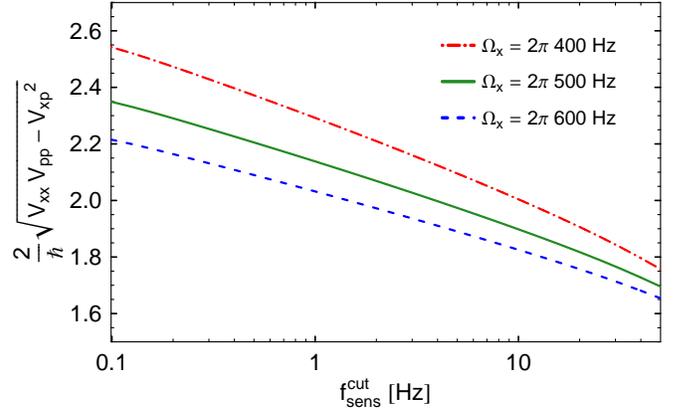}
\caption{Uncertainty product of the conditional state versus
sensing noise cut-off frequency for three different sensing noise
levels. Force noise contributions are the same as in
Fig.\,\ref{fig:cuttoff1}. The second-order moments formally
diverge when downshifting the cut-off frequency.}
\label{fig:cuttoff2}
\end{figure}

It should be emphasized that the conditional second-order moments
can diverge, if the sensing noise rises towards low frequencies,
and therefore a cut-off frequency must be chosen carefully. This
issue is illustrated by Figs.\,\ref{fig:cuttoff2} where the
cut-off frequency of the sensing noise is varied, while the
classical force noise contributions are held fixed. This
divergence can be explained as follows: for a free mass, the
effect of radiation pressure noise diverges towards low
frequencies.
Hence the boundedness of the conditional variances depends
crucially on the the motion of the test mass at low frequencies.
Furthermore the mirror thermal noise with $S(f) \sim 1/f$ formally
leads to a logarithmic divergence. In the case of a real
experimental set-up the mirrors are suspended as pendulums and the
mirror thermal noise should also exhibit a low-frequency cut-off
--- but more importantly the low frequency noise will be canceled
out in a subsequent verification stage, as it will be shown in a
forthcoming paper~\cite{CDMMRSS2008}. Such a cancelation arises
from the fact that for the mirror thermal noise with frequencies
lower than the inverse of the sum of the preparation-stage and
verification-stage measurement time scales, their contribution to
errors in the preparation and verification measurements are the
same, and therefore cancels out, when the two sets of data are
compared with each other. This argument justifies an increase of
the cut-off frequency to a level of around 3\,Hz.

\section{Advanced LIGO configurations}
\label{Sec:AdvLIGO}

In this section we will investigate the performance of the planned
Advanced LIGO detector~\cite{advLIGO}, a second generation
gravitational-wave observatory, towards the preparation of
test-mass quantum states. It is planned that this large-scale
laser interferometer (cf. Fig.\,\ref{Fig:AdvLIGO}) --- 4\,km long
arm cavities consisting of 40\,kg mirrors --- starts its operation
in 2014. It will be nearly quantum noise limited in most of its
frequency band (10\,Hz to 10\,kHz), and will operate near or at
its SQL. Our previous investigations in Sec.\,\ref{Sec:Marko} and
\ref{Sec:NonMarko} have suggested that such a SQL-sensitivity
allows to prepare nearly Heisenberg-limited quantum states of
macroscopic test masses. Note that we will consider the
differential mode of the interferometer's four movable arm-cavity
mirrors which is equivalent to a single movable mirror in a single
detuned cavity~\cite{BuCh2003} --- with one quarter the mass of
each individual mirror, i.e. 10\,kg.

The classical noise budget of the Advanced LIGO detector has been
estimated by the simulation tool \emph{Bench}~\cite{bench}.
\begin{figure}[t]
\begin{tabular}{c}
\includegraphics[height=0.45\textwidth,angle=-90]{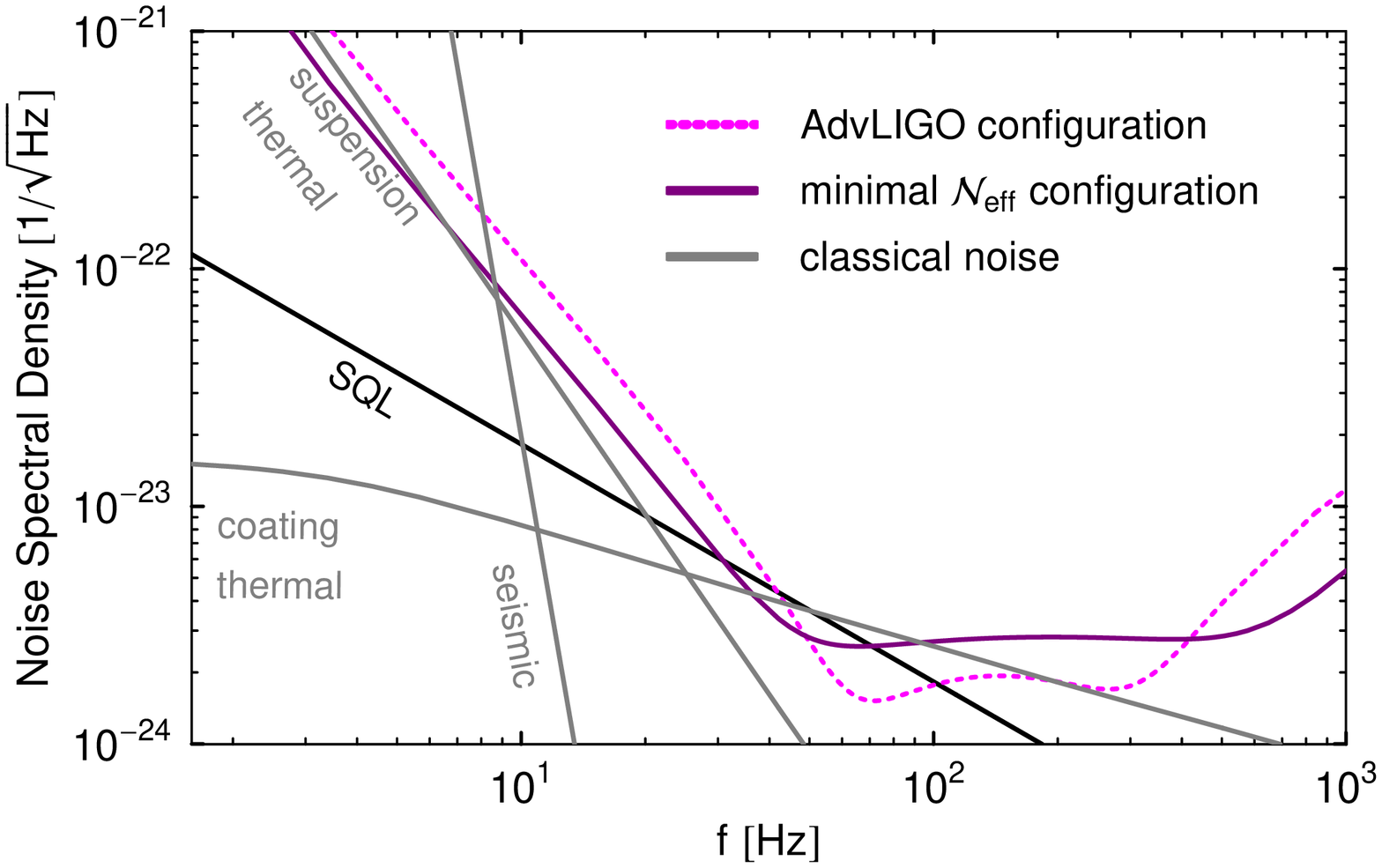}  \\
\includegraphics[height=0.4\textwidth]{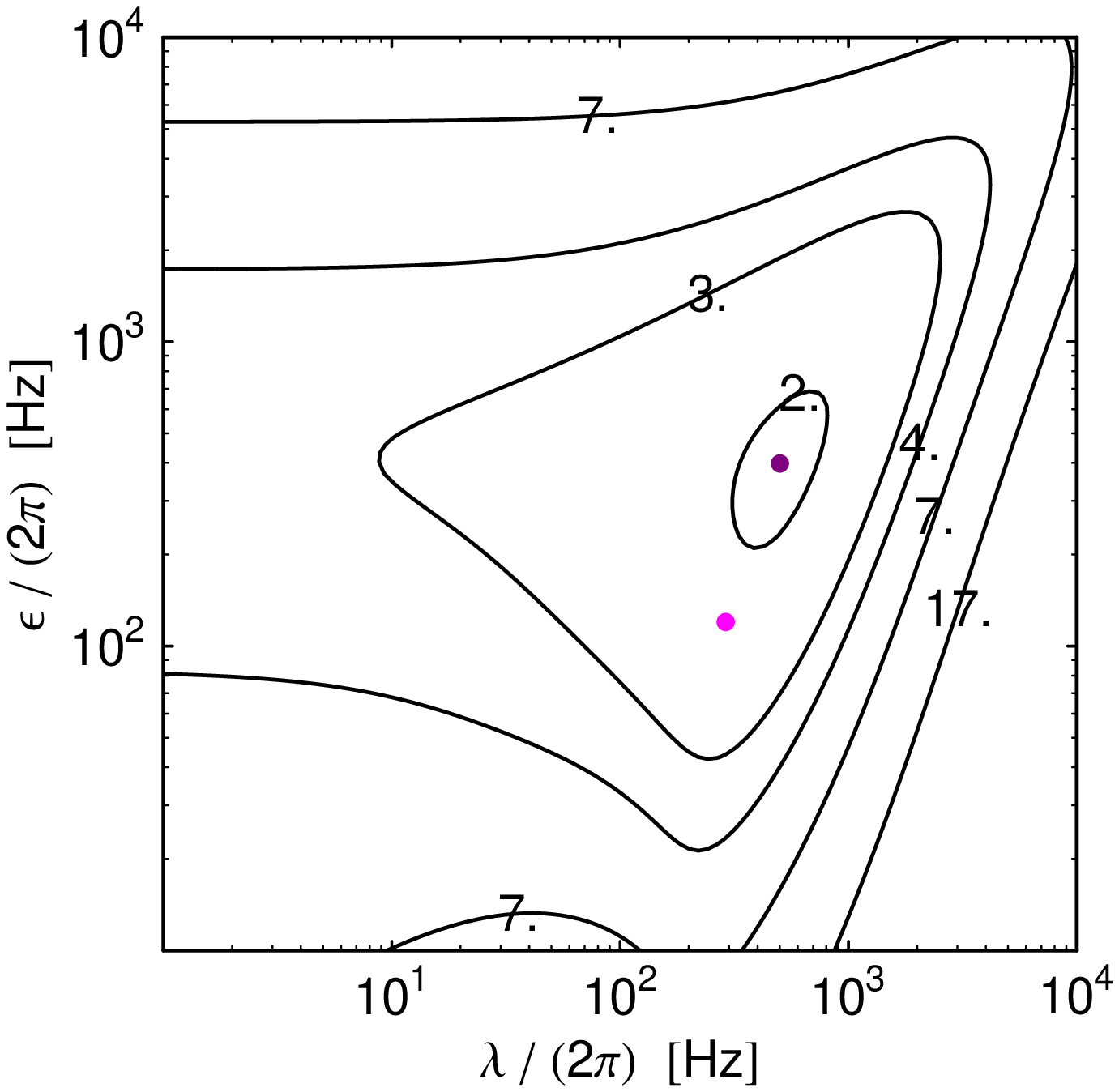}
\end{tabular}
\caption{\emph{Upper panel}: spectral densities of main noise
sources present in the Advanced LIGO detector: seismic, suspension
thermal and internal thermal noise (gray lines) as well as two
examples of the quantum noise (magenta and purple lines).
Pre-estimated classical non-Markovian noise budget is fitted by
rational functions with characteristic spectra as in
Fig.\,\ref{fig:cuttoff1}. \emph{Lower panel}: effective occupation
number $\mathcal{N}_{\rm eff}$ of conditional state (of the
differential mirror mode) versus effective detuning $\lambda$ and
effective bandwidth $\epsilon$. The dots mark the Advanced LIGO
broadband configuration state (magenta) and lowest occupation
number state (purple) when $\zeta =
0.7\,\pi$.}\label{fig:advligostateprep}
\end{figure}
We choose the same type of spectra as for the example
configuration in Sec.~\ref{SubSec:nonmarkonoise}, i.e. with
identical power laws and cut-offs in frequency, and adjust the
parameters such that the predicted Advanced LIGO classical noise
budget is well approximated. In contrast to
Sec.~\ref{SubSec:nonmarkonoise}, Advanced LIGO comprises finite
bandwidth cavities, which gives rise to a non-Markovian quantum
noise. Moreover, the detuned signal-recycling technique introduces
even the optical spring into the dynamics of the mirrors just as
in the case of a detuned cavity (cf. Sec.~\ref{SubSec:FinCav}). As
an example, the quantum noise of the Advanced LIGO broadband
configuration, which is optimal for the detection of neuron star
binary inspirals, is plotted in the upper panel of
Fig.\,\ref{fig:advligostateprep}.

We have carried out a full parameter search over the space of
signal-recycling parameters in order to optimize the configuration
with respect to the uncertainty product of the conditional state.
We have fixed the characteristic frequency $\iota_c$ (as defined
in Eq.\,(20) of Ref.~\cite{BuCh2003}) of the system, which is
basically determined by the fixed circulating optical power of
800\,kW which in turn determines the measurement frequency.
Furthermore, we have only considered a homodyne detection at the
Advanced LIGO broadband configuration quadrature, i.e. $\zeta =
0.7\,\pi$. The result of this optimization is shown in the lower
panel of Fig.\,\ref{fig:advligostateprep} which depicts the
effective occupation number $\mathcal{N}_{\rm eff}$ --- as
introduced in the Appendix\,\ref{Sec:Neff} --- versus the
effective detuning $\lambda$ and the effective bandwidth
$\epsilon$ --- these two quantities are defined in Eq.\,(18) of
Ref.~\cite{BuCh2003}. It clearly shows that the purity of the
conditional state benefits from a restoring optical spring, i.e. a
positive detuning facilitates the preparation of macroscopic
quantum states as it was shown before. Note that increasing the
effective bandwidth $\epsilon$ always gives rise to an additional
improvement, which has also been clarified before. The Advanced
LIGO broadband configuration with $\lambda = 2\pi\,290\,{\rm Hz}$
and $\epsilon = 2\pi\,120\,{\rm Hz}$ is marked with a dot in the
lower panel of Fig.\,\ref{fig:advligostateprep} and gives
$\mathcal{N}_{\rm eff} \approx 2.2$ while the other dot marks the
purest state at $\lambda = 2\pi\,500\,{\rm Hz}$ and $\epsilon =
2\pi\,400\,{\rm Hz}$ which gives $\mathcal{N}_{\rm eff} \approx
1.9$. An additional optimization of the homodyne detection angle
decreases this number only marginal.

Aside from the currently estimated classical noise budget, a more
optimistic scenario~\cite{ADHIKARI} has been investigated, in
which the seismic and suspension thermal noise are reduced by a
factor of ten, while the coating thermal noise is lowered by a
factor of three (in amplitude). Here the cut-off frequencies
remain the same.
\begin{figure}
\begin{tabular}{c}
\includegraphics[height=0.45\textwidth,angle=-90]{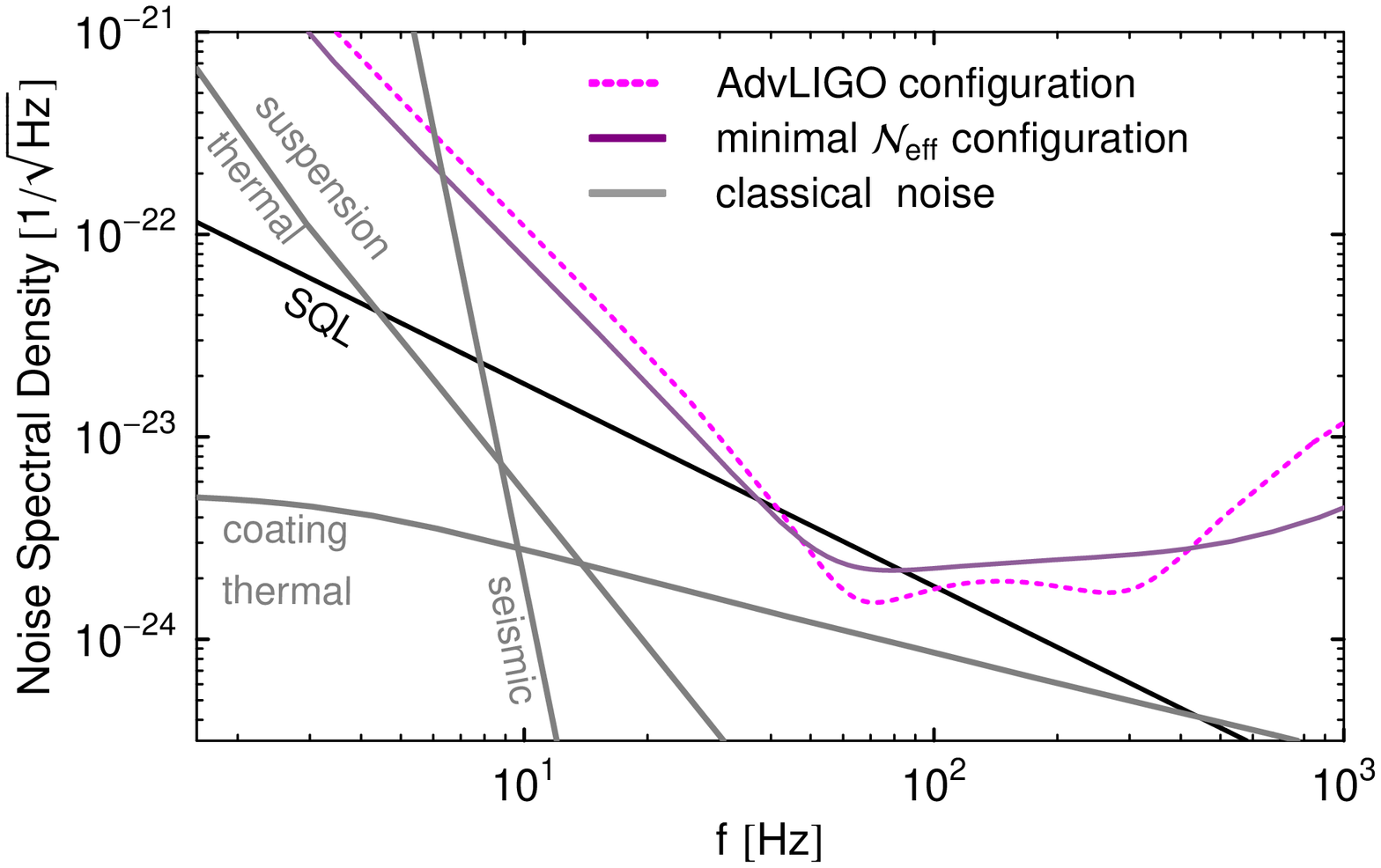}  \\
\includegraphics[height=0.4\textwidth,angle=-90]{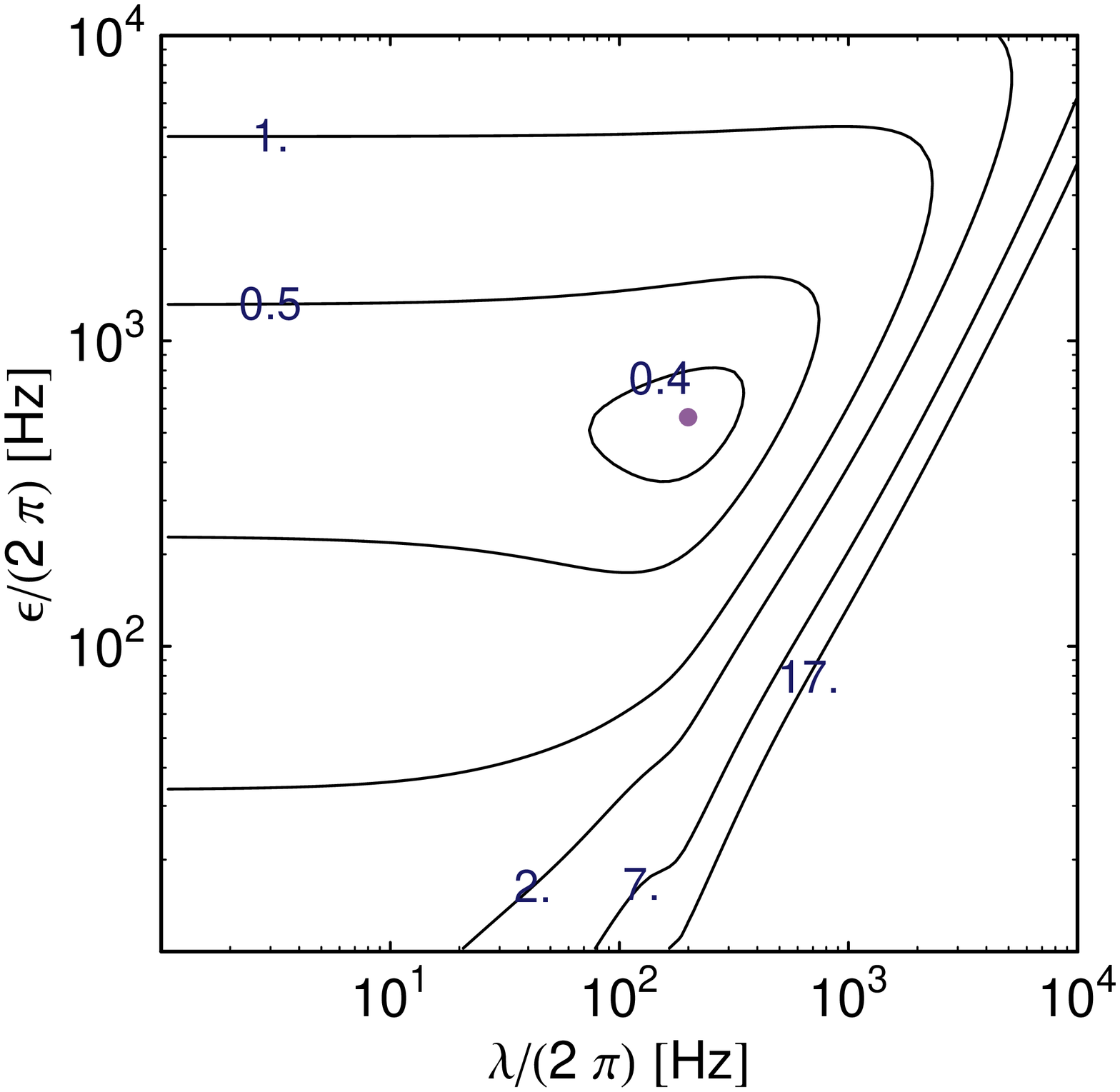}
\end{tabular}
\caption{\emph{Upper panel}: spectral densities of main noise
sources present in an improved Advanced LIGO detector. \emph{Lower
panel}: effective occupation number $\mathcal{N}_{\rm eff}$ of the
differential mirror mode's conditional state versus effective
detuning $\lambda$ and effective bandwidth $\epsilon$ for phase
quadrature detection. The dot marks the lowest occupation number
state.}\label{fig:advligostatepreplow}
\end{figure}
A rough optimization has revealed that the minimal achievable
effective occupation number drops down to $\mathcal{N}_{\rm eff}
\approx 0.38$ (cf. Fig.\,\ref{fig:advligostatepreplow}) for phase
quadrature detection. Further major improvements regarding the
classical noise level in the Advanced LIGO detector --- especially
the laser noise --- will even allow to entangle the cavity mirrors
in the north with those in the east arm.

\section{Conclusion}
\label{Sec:Con}

This paper has been devoted in great detail to a survey of the
first principles in the preparation of macroscopic Gaussian
quantum state of non-massless objects. We have motivated and
introduced the Wiener filter method in this context --- as an
advantage over the stochastic master equations --- and have given
a simple analytical expression for the covariance matrix of a
system under any continuous linear Markovian measurement process.
We have shown that in absence of any additional noise, the
conditional state is totally determined by the measurement noise.
Moreover, the purity of the conditional state is even equal to the
purity of the underlying measurement process (cf.
Eq.\,\eqref{Eq:conddet}). This provides an important insight into
the understanding of conditional states which was probably not
communicated before.

In Markovian measurements with non-correlated shot and
radiation-pressure noise, we have shown that the effective
occupation number of the conditional state is connected to the
factor the device beats the SQL, i.e. to the bandwidth within
which the classical noise is below the SQL (cf.
Eq.\,\eqref{Eq:SQLbeating} and Eq.\,\eqref{Eq:Neffcond}). For
$S_{\rm cl} < S_{\rm SQL}$ around the frequency $\Omega_{\rm cl}$,
at which the two classical noise spectra intersect, we find
$\mathcal{N}_{\rm eff} < 1/2$. We have shown that neither a
balanced homodyne detection of a non-phase quadrature nor
input-squeezing would help to get a more pure state
--- but they can significantly steer the shape of the conditional
state, e.g. the test-mass squeezing.

Furthermore, we have motivated that a simple power-recycled
Michelson interferometer is the ideal device to prepare
macroscopic entanglement~\cite{MRSDC2007}. We have shown that the
existence of entanglement in position and momentum between the two
end mirrors is closely related to the factor at which the
classical noise beats the SQL: a quantum measurement with a
flexible but frequency-independent homodyne detection angle and no
restriction to the optical power as an example, theoretically
requires the classical noise to be at least a factor of 1.5 below
the free-mass SQL at a certain sideband frequency.

Moreover, we studied mirror quantum-state preparation in
non-Markovian quantum-measurement systems. In the first instance
we have considered the conditional quantum state of a test mass
inside a finite-bandwidth system. It has been demonstrated that
even a quantum noise limited configuration does not allow the
preparation of a minimum Heisenberg uncertainty state --- due to
quantum entanglement between the test mass and the cavity mode,
which has a non-zero lifetime.

It has been pointed out that the purity of a conditional quantum
state of macroscopic test masses can benefit from introducing an
optical spring. This has been verified numerically for the quantum
noise limited regime.

Furthermore we have optimized the effective occupation number of
the differential mode of the planned Advanced LIGO
gravitational-wave detector in the presence of pre-estimated
realistic decoherence processes. It has been confirmed that
already a moderately reduced classical noise budget, such as for
an improved Advanced LIGO detector, allows us to prepare a nearly
pure quantum state of the mechanical mode under consideration.

Finally, we have explored the effective occupation number
achievable by Advanced LIGO, on the differential mode of its four
macroscopic test masses. It has been shown that an occupation
number of $\approx 2.2$ is readily reachable by the baseline
design, a moderate shift in optical parameters can achieve
$\approx 1.9$, while a moderate enhancement in classical noise
budget could achieve $\approx 0.38$. Third-generation
gravitational-wave detectors, or prototype interferometers
specifically designed for testing macroscopic quantum mechanics
would be able to surpass this moderate enhancement of Advanced
LIGO, and reach deep into the quantum regime.

\begin{acknowledgments}
We thank all the members of the AEI-Caltech-MIT MQM discussion
group for very useful discussions. We thank K.S.~Thorne for
initiating this research project, and V.B.~Braginsky for important
critical comments. Research of H.M.-E., K.S. and Y.C. is supported
by the Alexander von Humboldt Foundation's Sofja Kovalevskaja
Programme. Y.C. and K.S. are also supported by the National
Science Foundation (NSF) grants PHY-0653653 and PHY-0601459, as
well as the David and Barbara Groce startup fund at Caltech.
Research of H.R. and R.S. is supported by the Deutsche
Forschungsgemeinschaft through the SFB No. 407. K.S. is also
supported by the Japan Society for the Promotion of Science
(JSPS). Y.M. is supported by NSF grant PHY-0601459, PHY-0653653,
NASA grant NNX07AH06G, NNG04GK98G and the Brinson Foundation.
Research of C.L.\, is supported by National Science Foundation
grants PHY-0099568 and PHY-0601459.
\end{acknowledgments}

\begin{appendix}

\section{Quantum Wiener filter} \label{Sec:QJ}

Here we directly evaluate the conditional generating functional
involving the observables $\hat x_l(t)$,
\begin{eqnarray}
J&\equiv& \mathrm{tr} \left[\hat{\rho}^{\rm cond}\ e^{{\rm i}
\sum_l \alpha_l \hat x_l(t)}
\right] \nonumber \\
&=& \mathrm{tr} \left[ \hat{\rho}\ \mathcal{P}_{[\hat y (t') = y
(t)\,|\,t' < t]}\ e^{{\rm i} \sum_l \alpha_l \hat x_l(t)}
\right]\,,
\end{eqnarray}
with
\begin{eqnarray}
-{\rm i} \ \frac{\partial}{\partial \alpha_l}|_{\alpha_l=0} \ J
&=& x_{l}^{\rm cond} (t) \\
(-{\rm i})^2 \ \frac{\partial^2}{\partial \alpha_l\, \partial
\alpha_m}|_{\alpha_l=\alpha_m=0} \ J &=& V_{lm}^{\rm cond}\,.
\end{eqnarray}
If we then write the projection operator as a path integral,
\begin{eqnarray}
\mathcal{P}_{[\hat y (t') = y (t)\,|\,t' < t]}\propto \int D[k]
e^{{\rm i} \int_{-\infty}^{t} dt' k(t') \left[\hat y(t') -
y(t')\right]}\,,
\end{eqnarray}
we have
\begin{equation}
J\propto \int D[k] \mathrm{tr} \left[ \hat{\rho}\ e^{{\rm i}
\sum_l \alpha_l \hat x_l(t)+{\rm i} \int_{-\infty}^{t} dt' k(t')
\left[\hat y(t') - y(t')\right]} \right]\,.
\end{equation}
For a Gaussian state $\hat \rho$ and any linear observable $\hat x
=\hat x^\dagger$, if $\mathrm{tr}[\hat \rho\ \hat x] =0$, then we
always have
\begin{equation}
\mathrm{tr} \left[\hat\rho \ e^{i\hat x}\right] = e^{-\langle \hat
x^2\rangle/2}\,.
\end{equation}
Using this property, we obtain
\begin{widetext}
\begin{equation}
J\propto \int D[k] e^{-\left\langle \left[\sum_l \alpha_l \hat
x_l(t) + \int_{-\infty}^{t} dt' k(t') \hat
y(t')\right]^2\right\rangle/2}  e^{- {\rm i} \int_{-\infty}^{t}
dt' k(t')   y(t')}
\end{equation}
Suppose  Eqs.\,(\ref{xldecompose}) and (\ref{Ryindependence})
hold, i.e.,
\begin{equation}
\hat x_l(t) = \hat R_l(t) +\int_{-\infty}^t dt' K_l(t-t') \hat
y(t')\,,\quad \langle \hat R_l(t) \hat y(t')\rangle =0\,,\;\forall
t'<t\,,
\end{equation}
then
\begin{eqnarray}
\left\langle \left[\sum_l \alpha_l \hat x_l(t) +
\int_{-\infty}^{t} dt' k(t') \hat y(t')\right]^2\right\rangle =
\sum_{lm} \alpha_l \alpha_m \langle \hat R_l(t) \hat R_m(t)\rangle
\nonumber + \int_{-\infty}^t  \int_{-\infty}^t  dt' dt'' \tilde
k(t') \tilde k(t'') \langle  \hat y(t') \hat y(t'')\rangle\,,
\end{eqnarray}
where we have defined
\begin{equation}
\tilde k(t') = k(t') + \sum_l \alpha_l K_l(t-t')\,.
\end{equation}
Using $\tilde k(t)$ as the new integration variable, $J$ can be
re-written as
\begin{equation}
J \propto \exp\left[ -\frac{1}{2} \sum_{lm} \alpha_l \alpha_m
\langle \hat R_l(t) \hat R_m(t)\rangle \right] \exp\left[ {\rm i}
\int_{-\infty}^t dt' \sum_l \alpha_l K_l(t-t')y(t')\right]\,,
\end{equation}
which justifies Eqs.~\eqref{xlc} and \eqref{Vlmc}.
\end{widetext}

\section{Effective occupation number} \label{Sec:Neff}

The uncertainty product of Gaussian states is a true measure of
the purity and therefore a reasonable measure of the quantum-ness
of the state. Trying to reconstruct, as commonly done, the number
of quanta, the so-called {\it occupation number}, may not always
be the most fundamental figure of merit: squeezed states, for
example, can have high occupation numbers, yet they should be
considered probably more quantum than vacuum states. Moreover, the
definition of an occupation number requires a well-defined,
real-valued eigenfrequency, which does not always naturally exist.

The uncertainty product can be converted back into an {\it
effective occupation number} by using the relation
\begin{equation}\label{Eq:Neff}
2\,\mathcal{N}_{\rm eff} = \frac{2}{\hbar}\,\sqrt{V_{xx}\, V_{pp}
- V_{xp}^2} - 1\,.
\end{equation}
If a state has no correlation in position and momentum this
effective occupation number should be interpreted as follows:
suppose that the variances in position and momentum are given and
produced by a perfect harmonic oscillator in a quadratic potential
having an arbitrary but real-valued eigenfrequency $\omega_{\rm
eff}$. Then the effective occupation number is obtained by
minimizing the total energy divided by the energy of each quanta
with respect to that eigenfrequency $\omega_{\rm eff}$. This
strategy reads
\begin{equation} \label{Eq:Neffrel}
\mathcal{N}_{\rm eff} = \min_{\omega_{\rm
eff}}\left\{\frac{1}{\hbar\,\omega_{\rm eff}}\left(\frac{V_{
pp}}{2\,m} + \frac{m\,\omega_{\rm eff}^2\,V_{
xx}}{2}\right)-\frac{1}{2}\right\}\,,
\end{equation}
where the minimum is achieved at
\begin{equation} \label{Eq:eigeneff}
\omega_{\rm eff} = \sqrt{\frac{V_{pp}}{m^2\,V_{xx}}}\,.
\end{equation}
Thus, the effective occupation number is the minimal occupation
number one could obtain when assuming to have a harmonic
oscillator with no correlation in position and momentum and an
effective eigenfrequency as given in Eq.\,\eqref{Eq:eigeneff}.

For a state with correlation in position and momentum, the
effective occupation number still gives the minimal occupation but
with respect to two other orthogonal quadratures which are not
position and momentum. Moreover, the effective occupation number
is an interesting quantity because it in fact determines the {\it
von~Neumann~entropy} of a state~\cite{Zurek} as given by
\begin{equation}
\mathcal{S} = (\mathcal{N}_{\rm eff}+1)\log(\mathcal{N}_{\rm
eff}+1) - \mathcal{N}_{\rm eff} \log \mathcal{N}_{\rm eff}\,.
\end{equation}

\end{appendix}

\end{document}